\pdfoutput=1
\documentclass[superscriptaddress,nofootinbib,prl,two column]{revtex4}
\usepackage{CJK}
\usepackage[utf8]{inputenc}
\usepackage[T1]{fontenc}

\usepackage[colorlinks=true, pdfstartview=FitV, linkcolor=blue, citecolor=blue,
urlcolor=blue]{hyperref}
\usepackage[dvipdfmx]{graphicx}
\usepackage{amsmath,amssymb}
\usepackage{graphicx}
\usepackage{epstopdf}
\usepackage{bbold}
\usepackage{comment}
\usepackage{soul}
\usepackage{lipsum}
\usepackage{gensymb}
\usepackage{xcolor}
\usepackage{slashed}
\usepackage{subfigure}
\usepackage{bm}
\usepackage{mathrsfs}
\usepackage{physics}

\usepackage[normalem]{ulem}

\begin{document}
\title{Identifying $\alpha$-cluster configurations in $^{20}$Ne via ultracentral Ne+Ne Collisions}

    \author{Pei Li}
	\affiliation{Key Laboratory of Nuclear Physics and Ion-beam Application (MOE), Institute of Modern Physics, Fudan University, Shanghai 200433, China}
	\affiliation{Shanghai Research Center for Theoretical Nuclear Physics, NSFC and Fudan University, Shanghai $200438$, China}
	\author{Bo Zhou}
    \email[]{zhou$_$bo@fudan.edu.cn}
	\affiliation{Key Laboratory of Nuclear Physics and Ion-beam Application (MOE), Institute of Modern Physics, Fudan University, Shanghai 200433, China}
	\affiliation{Shanghai Research Center for Theoretical Nuclear Physics, NSFC and Fudan University, Shanghai $200438$, China}
	\author{Guo-Liang Ma}
	\email[]{glma@fudan.edu.cn}
	\affiliation{Key Laboratory of Nuclear Physics and Ion-beam Application (MOE), Institute of Modern Physics, Fudan University, Shanghai 200433, China}
	\affiliation{Shanghai Research Center for Theoretical Nuclear Physics, NSFC and Fudan University, Shanghai $200438$, China}

\begin{abstract}

The initial-state geometry in relativistic heavy-ion collisions provides a novel probe to nuclear cluster structure. For $^{20}$Ne, a novel approach is proposed to distinguish between the cluster configurations (5$\alpha$ versus $\alpha + ^{16}$O) in order to gain insight into nuclear structure transitions governed by many-body quantum correlations. Through analytical calculations with the microscopic Brink model and event-by-event simulations using the hydrodynamic framework, we establish the normalized symmetric cumulant NSC (3, 2) and the Pearson coefficient $\rho_2 (v_{2}^{2},\ \delta [p_{\mathrm{T}}])$ as quantitative discriminators to reveal enhanced cluster degrees of freedom in the ground state of $^{20}$Ne. The ultracentral Ne+Ne collisions at the LHC can experimentally identify these two competing configurations via these flow correlation observables, opening a new paradigm for probing clustering in light nuclei.
\end{abstract}

\maketitle

\paragraph*{Introduction.}
Understanding the cluster structure of light nuclei (A$\leq$20) plays a crucial role in many-body physics such as Bose-Einstein condensation (BEC)~\citep{Anderson:1995gf,Schunck:2008na} and nucleosynthesis processes in astrophysics~\citep{Freer:2014qoa,Chernykh:2007zz,Kahl:2023fphy}. In particular, $\alpha$-like four-nucleon correlations are intriguing due to their property of cluster geometric arrangement substructures~\citep{Gamow1930MassDC,10.1143/PTPS.E68.464}. In complex cluster configurations, the interplay of the shell model, collective motion, and single-particle degrees of freedom presents a fundamental challenge due to the combinatorial explosion of inter-cluster dynamical variables. This strongly correlated quantum many-body problem leads to computational complexity when modeling cluster structures in heavily deformed nuclei, where the dimensionality of the configuration space grows exponentially with nucleon number~\cite{nolting2009fundamentals}.

Unlike conventional scattering experiments at low energies, ultra-relativistic heavy-ion collisions provide access to nuclear structure through flow correlation observables, imprinting their intrinsic structures onto the initial geometry of quark-gluon plasma (QGP)~\citep{ALICE:2018lao,Zhang:2021kxj}. The transverse momentum spectra $\mathrm{d}N/\mathrm{d}p_{\mathrm{T}}$ of final hadrons reflect the nature of the final state of the QGP evolution, analyzed in terms of a Fourier expansion in azimuthal angle $\phi$: $\frac{\mathrm{d}N}{\mathrm{d}p_{\mathrm{T}}\mathrm{d}\phi} \propto 1+2\sum_{n=2}v_n(p_{\mathrm{T}})\cos n(\phi-\Phi_n)$. The flow coefficients $v_n$, related to the spatial eccentricity $\varepsilon_{n}$ of the initial state, encode the cluster structures of colliding nuclei, which then manifest as final-state momentum anisotropies~\cite{Ollitrault:1992bk,Heinz:2013th,Joslin10101983,PhysRevC.83.064904,Moreland:2014oya,YuanyuanWang:2024sgp}. For heavy nuclei, deformation descriptions including quadrupole and triaxial deformation have been extensively validated through U+U collisions~\cite{Zhang:2024ake} at STAR experiment and Xe+Xe collisions~\cite{Zhao:2024lpc,ALICE:2021gxt} at the CERN-LHC experiment, with clear manifestations in final-state observables. While for light nuclei, the deformation framework fails to capture essential physics and necessitates the adoption of cluster structure descriptions. The recently conducted O+O and Ne+Ne experiments at the LHC~\cite{ATLAS:2025nnt,ALICE:2025luc,CMS:2025tga} are providing suitable experimental platforms for studying complex cluster structures. Their high-precision ultracentral measurements provide the feasibility of accurately detecting the cluster structure in light nuclei, demanding urgent theoretical inputs.

As a touchstone for validating nuclear structure models, the $^{20}$Ne nucleus has a transition feature from mean-field to cluster structure~\cite{Zhou:2019cjz,Zhou:2023vgv,10.1143/PTPS.68.303,PhysRevC.50.795,PhysRevC.52.1840,PhysRevC.83.014302} and it provides a unique opportunity to investigate the interplay between nuclear potential depth, single-nucleon orbital localization, and emergent clustering phenomena, offering critical insights into the quantum-liquid-to-cluster phase transition in finite nuclei~\cite{ebran_how_2012}. The calculations with Antisymmetrized Molecular Dynamics (AMD) model indicate that bowling-pin-like $\alpha+^{16}$O is a major configuration of the ground-state rotational band of $^{20}$Ne to explain the existence of $K_\pi=0_1^\pm$ rotational band~\cite{10.1143/PTP.40.277}. The cluster structure of $^{20}$Ne gradually changes to the shell-model structure with increasing angular momentum~\cite{10.1143/ptp/93.1.115}. From the perspective of the nonlocalized THSR model~\cite{tohsaki_alpha_2001,zhou_nonlocalized_2013,Zhou:2019cjz,Horiuchi:2017cjn,PhysRevC.89.034319}, two clusters are bound together to make nonlocalized motion. There is no obvious parameter to limit the two clusters from forming a certain deformation. Recent theoretical investigations within the Algebraic Collective Model (ACM) have revitalized the hypothesis of the bi-pyramidal configuration in $^{20}$Ne associated with $\mathcal{D}_{3h}$ symmetry (comprising 5$\alpha$ clusters arranged in a bi-pyramid)~\cite{Bijker:2020gbl}. On the other hand, from the view of the simple nuclear shell model, the ground state of the $^{20}$Ne nucleus tends to exhibit a spherical configuration. Therefore, understanding the competing cluster configurations (e.g., $5\alpha$ vs. $\alpha$+$^{16}$O) in $^{20}$Ne is critical to unravelling how nuclear structure models transition between mean field and cluster-dominated regimes. However, experimental investigations through traditional low-energy reactions remain challenging due to limitations such as insufficient sensitivity to cluster configurations and difficulties in isolating reaction pathways.

In theoretical studies in high energy heavy-ion collisions, the linear relations exist $v_n=\kappa_n\varepsilon_n$~\cite{Alver:2010gr,PhysRevC.93.024907,PhysRevC.86.044908,Nijs:2020ors,Bernhard:2019bmu}, where the response coefficients $\kappa_n$ reflect the transport properties of the QGP. On the other hand, the inverse transverse size $d_{\perp}$ is typically considered to reflect the nature of the final-state mean transverse momentum $[p_{T}]$~\cite{Bozek:2012fw,Schenke:2020uqq}. Under different transverse planes corresponding to various Euler rotation angles, the correlated fluctuations between different $\varepsilon_{n}$ components $\langle\varepsilon_{n}^{2}\varepsilon_{m}^{2}\rangle$~\cite{Bilandzic:2010jr,Bilandzic:2013kga,ALICE:2021adw} or the mixed correlated fluctuations with the transverse plane size $\langle \varepsilon_{n}^{2}\delta d_{\perp} \rangle$~\cite{Bozek:2016yoj,Jia:2021qyu} would be very significant observables for the initial geometry of the QGP~\cite{Behera:2023nwj,Zhang:2024vkh,ATLAS:2022dov}. To clarify the cluster configuration problem of $^{20}$Ne, we can establish the connection between the cluster wave function and the initial state of the QGP in heavy-ion collisions, through a reasonable cluster model (i.e., the microscopic Brink wave function~\cite{PhysRev.59.37,10.1143/PTPS.62.90,Zhongzhou:2017mbd} or the nonlocalized THSR wave function). We utilize the properties of the eccentricity and the inverse transverse size to investigate the cluster structure of light nuclei described with the wave function.

This Letter introduces analytical and numerical nuclear imaging to resolve geometric cluster structures at the heavy-ion collision instant. Unlike conventional flow observables of heavy-ion collisions in $^{20}$Ne~\cite{Giacalone:2024luz,Giacalone:2024ixe}, which are effective for characterizing shape deformation but fundamentally lack the resolving power to discriminate between competing configurations ($\alpha+^{16}$O and $5\alpha$), we identify two discriminative observables: normalized symmetric cumulants NSC$(3,\ 2)$ and Pearson coefficients $\rho_2(v_2^2,\delta [p_{\mathrm{T}}])$. Validated through the localized Brink cluster model and event-by-event simulations using the hydrodynamic framework, these two observables establish the first quantitative constraints for emergent cluster structures in light nuclei via heavy-ion collisions, revealing previously inaccessible many-body quantum correlations.

\paragraph*{Microscopic Brink cluster model.}
The Brink cluster model~\cite{brink_alphaparticle_1966} is a widely used and important microscopic theoretical approach to deal with cluster structures of light nuclei. The Brink cluster model effectively captures the antisymmetrization effects of nucleons and their spatial correlations. Furthermore, through Generator Coordinate Method (GCM) calculations, multi-cluster systems can be accurately described in light nuclei~\cite{Zhou:2019cjz}.
The Brink cluster wave function $\Phi^{\text{B}}$ for 5$\alpha$ clusters in $^{20}$Ne is defined as, 
\begin{equation}
\begin{aligned}
\Phi^{\text{B}} (\bm{R}_1,\bm{R}_2&,\bm{R}_3,\bm{R}_4,\bm{R}_5) =
\\&\frac{1}{\sqrt{20!}} \mathcal{A} [\phi_1(\bm{R}_1)...\phi_5(\bm{R}_2)...\phi_{20}(\bm{R}_5) ].
\label{Eq:1}
\end{aligned}
\end{equation}
With the single-nucleon wave function, 
\begin{align}
\phi_i(\bm{R}_k) = (\frac{1}{\pi b^2})^{\frac{3}{4}}\ \text{exp}\left[ {-\frac{1}{2b^2}(\bm{r}_i-\bm{R}_k)^2} \right] \chi_i \tau_i.
\label{Eq:2}
\end{align}
Here, $\phi_i(\bm{R}_k)$ is the single-nucleon wave function characterized by the Gaussian center parameter $\{\bm{R}_k\}$ and harmonic oscillator size parameter $b$. The generator coordinates \(\bm{R} \equiv\{R_1, R_2, R_3, R_4, R_5\} \) parameterize the spatial configurations of the 5$\alpha$ clusters in the \(^{20}\mathrm{Ne}\) nucleus. The $\chi_i$ and $\tau_i$ components are the spin and isospin parts, respectively.

Although the single Brink cluster model struggles to accurately describe the ground state of \(^{20}\mathrm{Ne}\), it has been demonstrated that the nonlocal THSR wave function provides a superior description of the excited low-lying states of \(^{20}\mathrm{Ne}\)~\cite{zhou_nonlocalized_2013}. However, in the single Brink cluster model, the spatial correlations among the 5$\alpha$ clusters can be explicitly expressed. As an approximation, this allows for a reasonable investigation of clustering-related effects in high-energy heavy-ion collisions. Within the Brink cluster model, the density distribution at position $\bm{a}$ can be defined as
\begin{align}
\rho(\boldsymbol{a})=\frac{\langle\Phi^{\text{B}}(\mathbf{R})|\frac{1}{20}\sum_{j=1}^{20}\delta(\bm{r}_j-\bm{r}_\mathrm{g}-\bm{a})|\Phi^{\text{B}}(\mathbf{R})\rangle}{\langle\Phi^{\text{B}}(\mathbf{R})|\Phi^{\text{B}}(\mathbf{R})\rangle}.
\label{Eq:3}
\end{align}

The integral part involving the center-of-mass coordinate $\bm{r}_{\mathrm{g}}=\frac{1}{20}\Sigma_{j=1}^{20}\bm{r}_{j}$ can be separated, that is, $\delta(\bm{r}_j-\bm{r}_{\mathrm{g}}-\bm{a}) \rightarrow \delta(\bm{r}_{\mathrm{g}})\delta(\bm{r}_j-\bm{a})$. The $\delta(\bm{r}_j-\bm{a})$ is a single-body operator while $\delta(\bm{r}_{\mathrm{g}})$ is an $A$-body operator and requires transforming the center-of-mass part into the momentum space for calculations. In this case, the nucleon density could be determined analytically by Eq.~(\ref{Eq:3}), which is critical for subsequent discussions on correlated observables. The initial conditions of the collision between two nuclei are uniquely defined by the nucleon distribution in the 2D-transverse plane $\rho(r_{\perp},\phi)=\int[\mathcal{R}_{p}(\phi_{p,1},\theta_p,\phi_{p,2})\rho(x,y,z)+\mathcal{R}_{t}(\phi_{t,1},\theta_t,\phi_{t,2})\rho(x,y,z)]/2\ \mathrm{d}z$, where $\mathcal{R}_{p}$ and $\mathcal{R}_{t}$ are the Euler rotation matrix for the projectile and target nuclei, respectively.

        \begin{figure*}[htb]
		\centering

        \subfigure
		{
			\begin{minipage}[b]{.29\linewidth}
				\centering
				\includegraphics[scale=0.36]{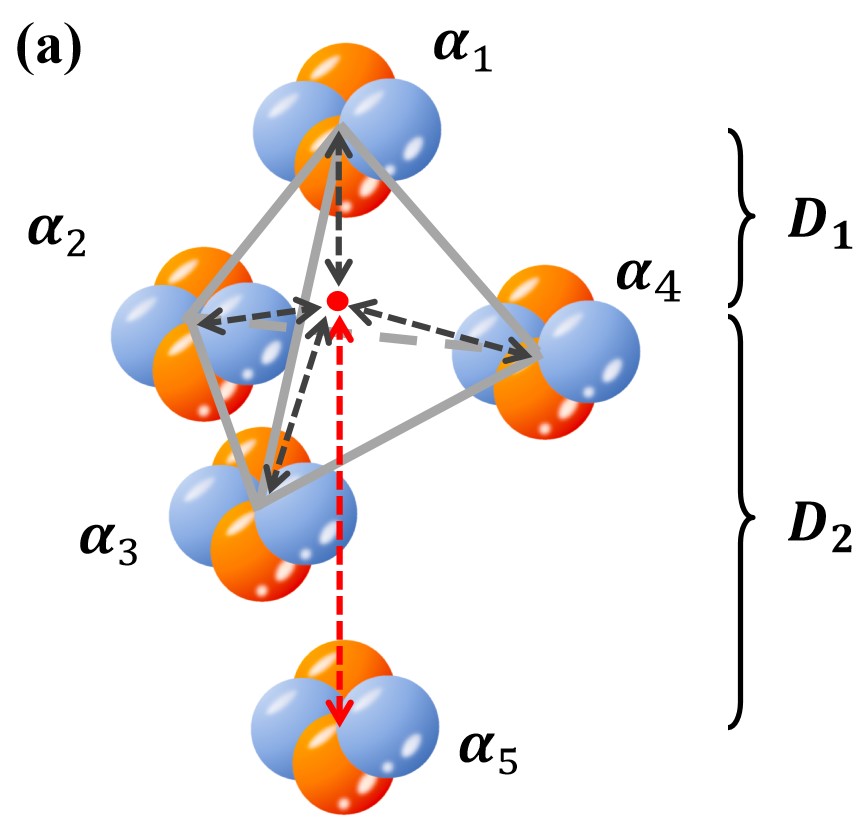}
			\end{minipage}
		}
        \subfigure
		{
			\begin{minipage}[b]{.33\linewidth}
				\centering
				\includegraphics[scale=0.39]{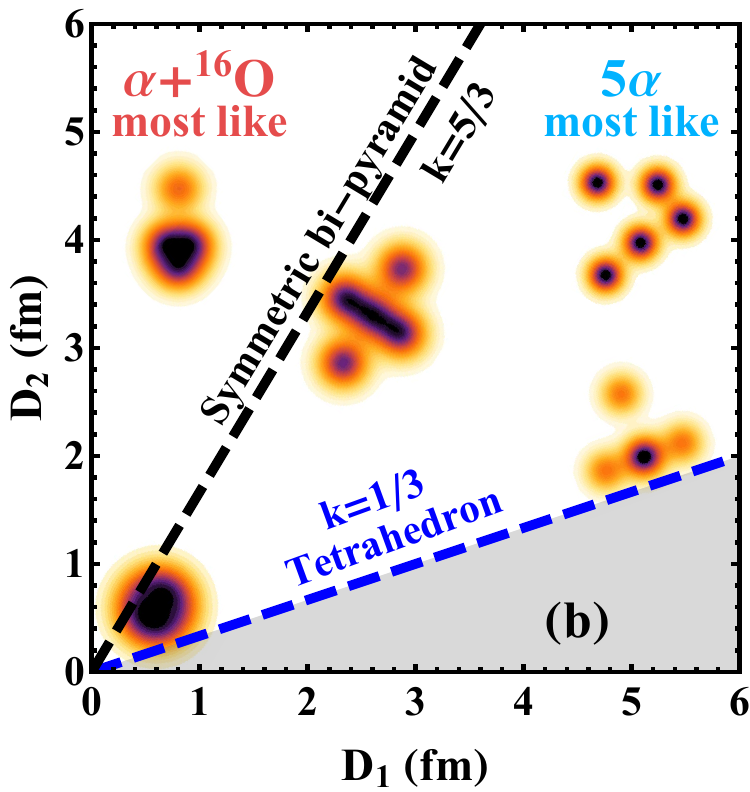}
			\end{minipage}
		}
        \subfigure
		{
			\begin{minipage}[b]{.33\linewidth}
				\centering
				\includegraphics[scale=0.545]{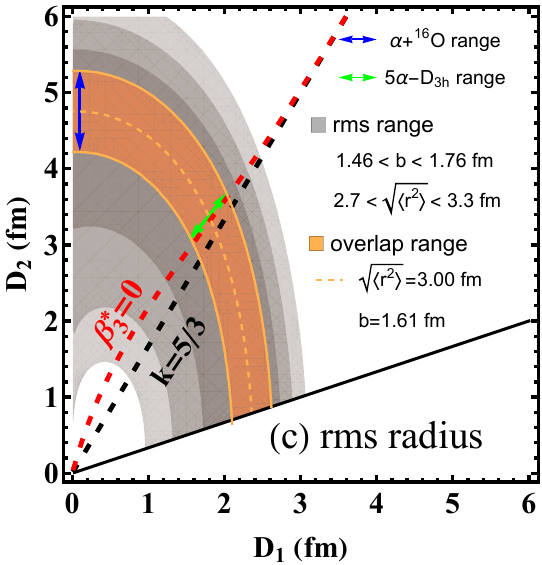}
			\end{minipage}
		}

    \caption{(Color online) (a) The 5$\alpha$ cluster structure diagram of $^{20}$Ne, where the four $\alpha$-particles $\alpha_{1\sim4}$ form a regular tetrahedral structure, and the parameter $D_1$ is the distance from each cluster to the center of the regular tetrahedron, while the parameter $D_2$ is the distance from the cluster $\alpha_{5}$ to the center of the regular tetrahedron. (b) The $D_1-D_2$ schematic diagram of several special structures of $^{20}$Ne with the ratio parameters $k=D_2/D_1$. (c) The schematic diagram for the selection range of the $D_1$, $D_2$ and $b$ for $^{20}\text{Ne}$. Each grey band represents the range of the rms radius within $\pm 10\%$ of the experimental value after fixing $b$ value. The orange band is the overlap region (OR) of all grey regions, which indicates the range of $D_1$ and $D_2$ that satisfies the constraint that the rms radius is within the range $2.7 < \sqrt{\langle r^2\rangle} < 3.3 \text{ fm}$ under the condition $1.46 < b < 1.76\text{ fm}$. The orange dashed line is the average configuration using $b = 1.61\text{ fm}$. The black dashed lines in (b) and (c) present the region of symmetric bi-pyramid 5$\alpha$ configuration satisfying $k = 5/3$.}
    \label{Fig:config}
    \end{figure*}
    
We consider $^{20}$Ne as the 5$\alpha$-cluster structure satisfying the bi-pyramidal configuration and introduce the structure characteristic length $D_1$ and $D_2$ to describe localized cluster structure, as shown in Fig.~\ref{Fig:config}(a). The parameter $D_1$ presents the distance from the $\alpha_1\sim\alpha_4$ cluster in the regular tetrahedron to the center of the tetrahedron, and the parameter $D_2$ presents the distance from another $\alpha_{5}$ cluster to the center of the regular tetrahedron. It should be noted that for the limit of the parameter $D_1\rightarrow0$, the 5$\alpha$ cluster structure degenerates into a two-cluster configuration of $\alpha+^{16}$O in which the $^{16}$O is a closed-shell and nearly perfect spherical structure~\cite{Yamaguchi:2023mya}. To facilitate the comparison of the structure of $^{20}$Ne, we define a structure parameter $k\equiv D_2/D_1$. For the same $k$ value, we only consider the shape of $^{20}$Ne and neglect the relative size of the nucleus. Under this definition, $^{20}$Ne can have several special shapes, as illustrated in Fig.~\ref{Fig:config}(b). The blue dashed line represents $k=1/3$, at which $^{20}$Ne adopts the tetrahedral structure with $\alpha_5$ lying on the same plane as $\alpha_2\sim \alpha_4$. The black dashed line represents $k=5/3$, at which $^{20}$Ne adopts the symmetric bi-pyramidal structure, $\alpha_1$ and $\alpha_5$ are symmetric with respect to the plane formed by $\alpha_2\sim \alpha_4$. Meanwhile, we have marked the regions with significant $\alpha+^{16}$O configuration features or $5\alpha$ configuration features, respectively.

For comparison, we chose suitable parameters for nuclear structures for $^{20}$Ne with the W-S distribution. The W-S distribution can be expressed more refinedly in the form of a three-parameter Fermi (3pF) model:
\begin{align}
\rho(r,\theta,\phi)=\rho_0(1+wr^2/R(\theta,\phi)^2)/[1+\exp(\frac{r-R(\theta,\phi)}{a})],
\label{Eq:4}
\end{align}
where $R(\theta,\phi)=R_0(1+\sum_{n ,m}\beta_n Y_{n,m}(\theta,\phi))$. The distribution is characterized by the size parameter $R_{0}=2.791 \ \mathrm{fm}$, the surface diffusion parameter $a =0.698 \ \mathrm{fm}$, the weight parameter $w =-0.168$~\cite{DEVRIES1987495} and the deformation parameters $\beta_n$. In the description of W-S distribution, derived under a sharp surface liquid-drop approximation, analytical estimation for ultracentral collisions reveals a quadratic dependence $\langle \varepsilon^2_{n} \rangle \propto a^\prime + b^\prime\beta_{n}^2$, where the parameters $a^\prime$ and $b^\prime$ depend on both the collision system and the collision energy.~\cite{Jia:2021qyu,Jia:2021tzt,Wang:2024ulq}. Therefore, we present the formula for calculating the deformation parameters here. For a given density distribution, we can perform theoretical calculations for the deformation parameters $\beta_n$ of the nuclear structure through the following integration in spherical coordinates~\cite{Ryssens:2023fkv,bookring,Scamps:2020fyu}:
\begin{align}
\beta_{n}^{*}=\frac{4\pi}{3A\langle r^{2}\rangle^{n/2}}\int\rho(\bm{r})r^{n}Y_{n,0}(\theta,\phi)d^{3}r,\ \ \ \ n\geq 2,
\label{Eq:5}
\end{align}
where $\langle r^{2} \rangle^{1/2}$ is root-mean-square (rms) radius. This provides an approximate conversion of the cluster model description into the traditional description using deformation parameters. Detailed information can be found in the Supplementary Material.

\begin{figure*}[htb]
		\centering

        \subfigure
		{
			\begin{minipage}[b]{.30\linewidth}
				\centering
				\includegraphics[scale=0.37]{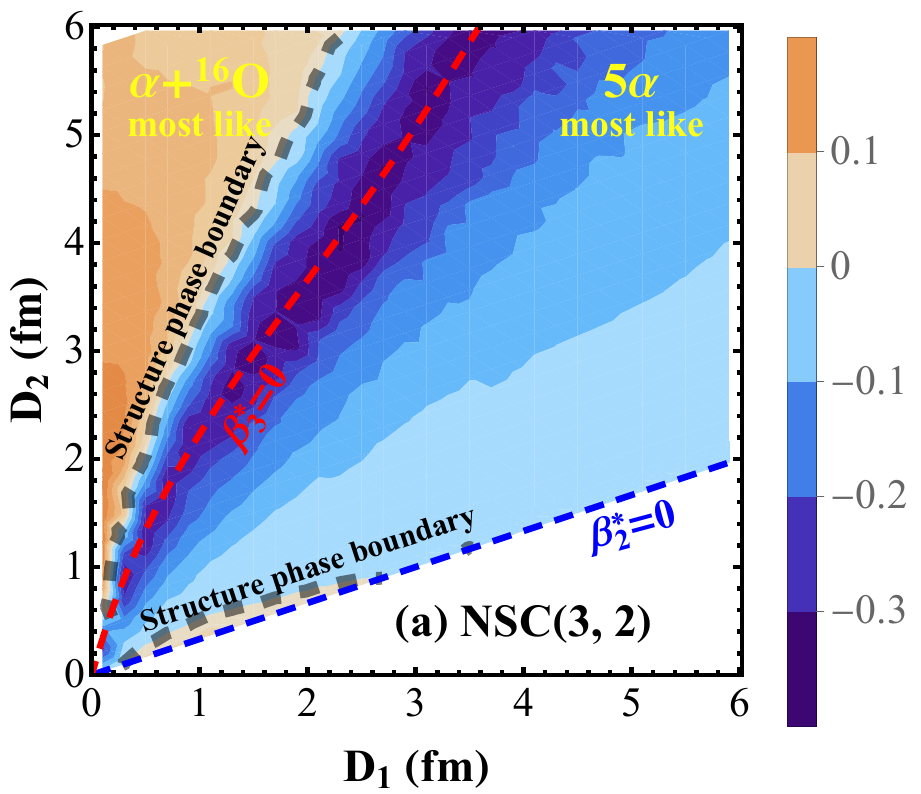}
			\end{minipage}
		}
        \subfigure
		{
			\begin{minipage}[b]{.30\linewidth}
				\centering
				\includegraphics[scale=0.37]{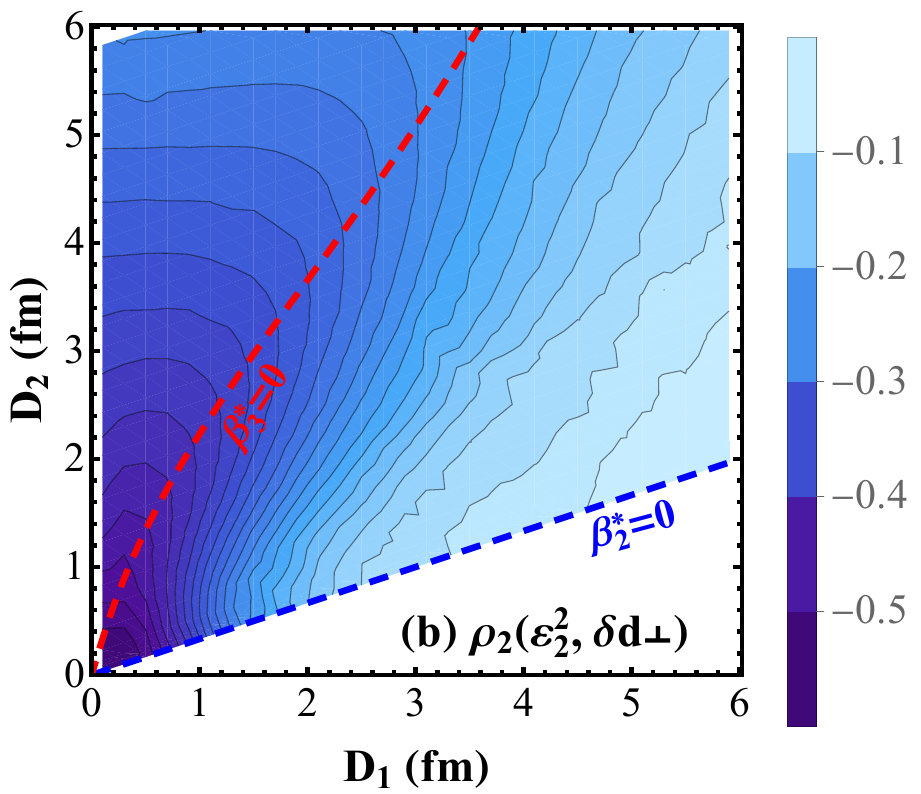}
			\end{minipage}
		}
        \subfigure
		{
			\begin{minipage}[b]{.30\linewidth}
				\centering
				\includegraphics[scale=0.37]{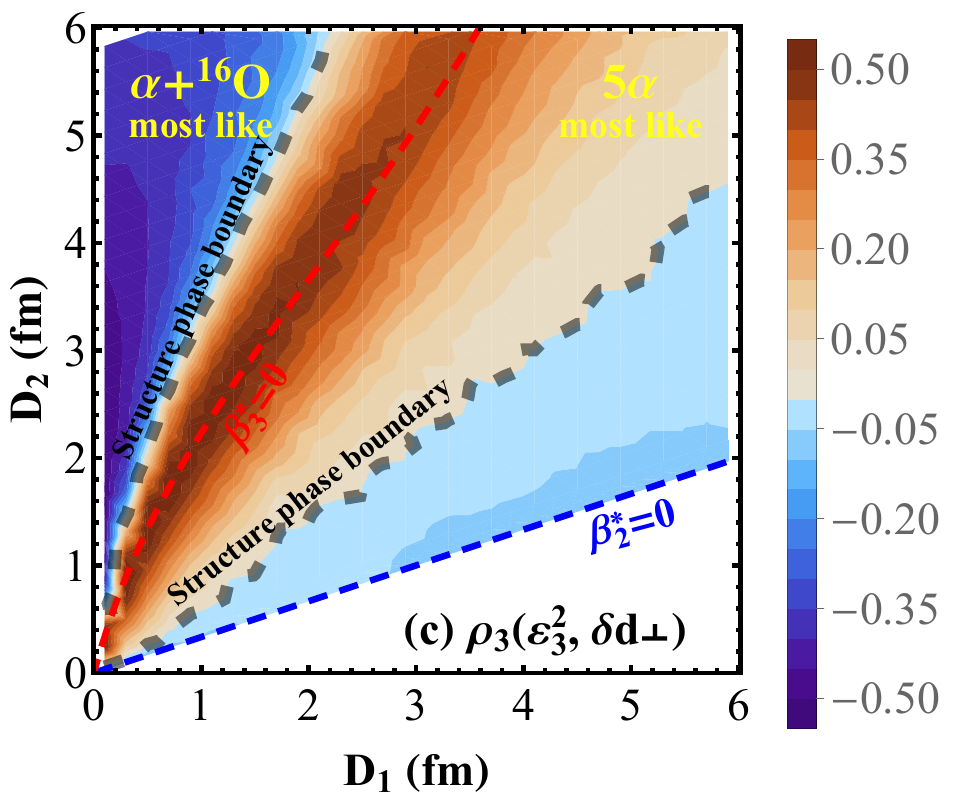}
			\end{minipage}
		}
        \subfigure
		{
			\begin{minipage}[b]{.30\linewidth}
            \hspace*{-0.92cm}
            	\centering
				\includegraphics[scale=0.52]{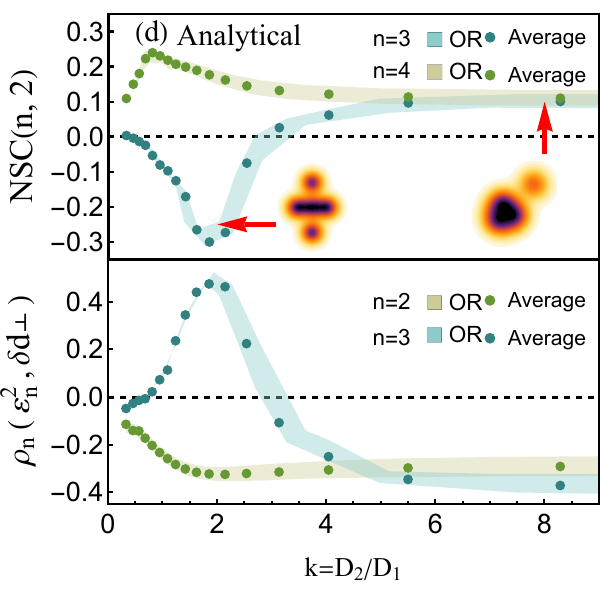}
			\end{minipage}
		}
        \subfigure
		{
			\begin{minipage}[b]{.30\linewidth}
                \hspace*{-0.92cm}
				\centering
				\includegraphics[scale=0.52]{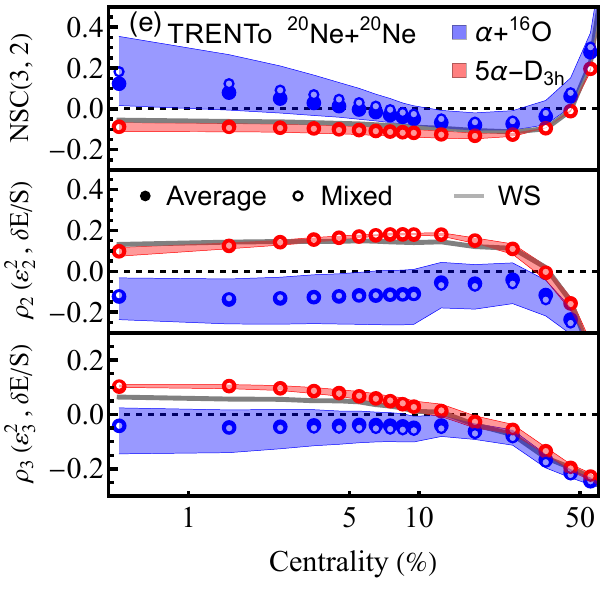}
			\end{minipage}
		}
        \subfigure
		{
			\begin{minipage}[b]{.30\linewidth}
                \hspace*{-0.92cm}
				\centering
				\includegraphics[scale=0.52]{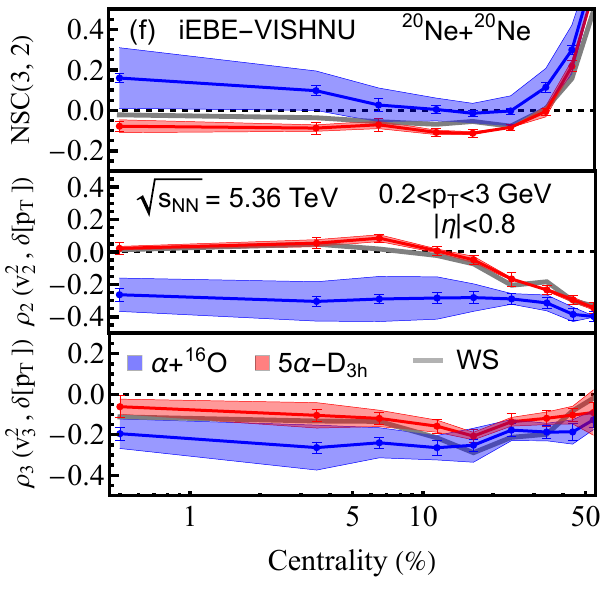}
			\end{minipage}
		}%
		
 \vspace{-1\baselineskip}
 
		\caption{(Color online) The analytical results on the dependence of the correlation observables on the parameters $D_1$ and $D_2$ with $b=1.76$ fm, (a): NSC$(3,\ 2)$, (b): $\rho_2$, (c): $\rho_3$. The blue and red dashed lines correspond to $\beta_{2}^{*}=0$ and $\beta_{3}^{*}=0$, respectively. The gray thick dashed lines are the structure phase boundaries separating the $\alpha$+$^{16}$O and $5\alpha$ configurations. (d): The analytical results on the dependences of NSC$(m,\ n)$ and $\rho_n$ on the ratio $k=D_2/D_1$ satisfying the rms radius requirements in Fig.~\ref{Fig:config}(c), the uncertainty is given by the systematic uncertainty of the analytical model. The initial-state results (e) and final-state results (f) with fluctuation effect using the T\raisebox{-0.5ex}{R}ENTo+VISHNU+UrQMD hydrodynamic framework are shown, where the shaded region represents the systematic uncertainty and the error bar indicates the statistical uncertainty.}
	\label{Fig:corr}
	\end{figure*}

\paragraph{Analytical Results.}
To account for the systematic uncertainty of the analytical theory and comprehensively cover the dual structural characteristics of $^{20}\text{Ne}$, we select the harmonic oscillator length parameter within $1.46<b<1.76 \text{ fm}$~\cite{Zhou:2019cjz,Zhou:2017jhz,10.1143/PTP.53.706}. To fit the experimental rms charge radius $\langle r^{2}\rangle^{1/2} \approx$ 3.00 fm~\cite{ANGELI201369} and determine a reasonable range for the structural parameters, we considered the $10\%$ error margin on the rms radius, which accounts for all plausible structural parameters for $^{20}\text{Ne}$ to the greatest extent possible. The 10\% variation guarantees that the lower bound of the $^{20}\text{Ne}$ radius exceeds the $^{16}\text{O}$ radius of 2.7 fm. In Fig.~\ref{Fig:config}(c), the orange band defines the selected $(D_1, D_2)$ parameter space under the $b$ constraint and represents the corresponding systematic uncertainty of the cluster model. The blue and green bands correspond to the parameter ranges for the $\alpha$+$^{16}$O and 5$\alpha$ configurations in our following hydrodynamic simulations, respectively, the spread of which contributes to the systematic uncertainty in our results. To ensure that the structure of $^{20}$Ne always adopts a bi-pyramidal configuration, we only retain the region where $k>1/3$.
    
The analytical results are obtained by considering the size of the clusters and neglecting the fluctuations in nucleon positions, thereby enabling complete isolation of the effects arising from the geometric configuration. Figure~\ref{Fig:corr}(a) shows the dependence of the normalized symmetric cumulant NSC$(3,\ 2)$ on the parameters $D_1$ and $D_2$. When $D_1$ is relatively small, NSC$(3,\ 2)$ is positive, corresponding to the $\alpha + \mathrm{^{16}O}$ configurations. While NSC$(3,\ 2)$ becomes negative as $D_1$ increases, corresponding to the $5\alpha$ configurations. There exists a structure phase boundary where NSC$(3,\ 2)=0$ with $k \approx 3$, and another structure phase boundary is the line with $k=1/3$ where $\varepsilon_{2}=0$. The correlation observable NSC$(3,\ 2)$ demonstrates pronounced sensitivity to nuclear structure effects induced by symmetry-driven geometric reorganization in many-body systems. Therefore, we can determine the configuration of $^{20}$Ne by the sign of the normalized symmetric values.

To investigate the properties of the Pearson correlation coefficients $\rho_n$, we modify the definition of the inverse area size in the 2D-transverse plane as 
\begin{align}
d_{\perp}=(\langle x^{2}y^{2}\rangle-\langle xy\rangle^{2})^{-1/4}.
\label{Eq:13}
\end{align}
The definition of $d_{\perp}$ involves the calculation of the average value of the fourth-order exponential $\langle x^{2}y^{2}\rangle$, which introduces more contribution from $\beta_{4}$. The detailed derivation process with the description of W-S distribution can be found in the Supplementary Material. This examination also demonstrates the consistency between the Brink cluster model and the W-S distribution, although the former is a boundaryless distribution and the latter has boundaries. Figures~\ref{Fig:corr}(b) and (c) present the dependence of the Pearson correlation coefficient $\rho_n$ on $D_1$ and $D_2$. The $\rho_n$ is influenced by $\beta_{2}^{*}$, $\beta_{3}^{*}$ and $\beta_{4}^{*}$ collectively, making the picture much more complex. The observed monotonic decrease of $\beta_2^{*}$ drives $\rho_2$ toward vanishing values, which is a key motivation for our redefinition of the inverse transverse size in the cluster model framework. Additionally, we observe that the NSC$(3,\ 2)$ and $\rho_3$ are not directly related to the absolute values of $D_1$ and $D_2$ but are related to the ratio $k$. To meet the rms radius requirements, we present the dependence of NSC$(n,\ 2)$ ($n$=3, 4) and $\rho_n$ ($n$=2, 3) on the ratio $k$ for the physical points on the orange region in Fig.~\ref{Fig:config}(c). Our analytical framework identifies NSC$(3,\ 2)$ and $\rho_3$ as candidate observables sensitive to the $^{20}$Ne configurations. The true performance of these observables must be validated through following hydrodynamic simulations, in which  $\rho_3$ fails to serve as a reliable discriminator in the final state owing to fluctuation effects.

\paragraph{Hydrodynamic Results.}
To incorporate the initial-state and final-state fluctuation effects during the QGP and hadronic evolution processes and to better match realistic experimental conditions, we simulate the full evolution of Ne+Ne collisions at $\sqrt{s_{NN}}=5.36\ \mathrm{TeV}$ using the (2+1)-dimensional viscous hydrodynamics framework T\raisebox{-0.5ex}{R}ENTo+VISHNU+UrQMD~\cite{ALICE:2024rpm,Zhao:2024lpc,YuanyuanWang:2024sgp}. The centrality is determined by the initial entropy density $S$. In the T\raisebox{-0.5ex}{R}ENTo model, the number of charged particles is given by $N_{ch}=\int dxdy\left[(T^p_A+T^p_B)/2\right]^{1/p}$, where $p=0$ in this work (we have checked that $p=1$ leads to identical conclusions). We use the multi-dimensional adaptive Monte Carlo (AMC) sampling method from Ref.~\cite{Li:2025src} to obtain the nucleon configurations from the cluster wave function, which are then used as input for the hydrodynamic framework. We characterize the inverse collision area size using the more physically relevant quantity $E/S$, derived from the initial-state energy density distribution, rather than $d_{\perp}$~\cite{Giacalone:2020dln}.

Figures~\ref{Fig:corr}(e) and (f) show the results from the initial-state T\raisebox{-0.5ex}{R}ENTo model and the final-state of hydrodynamic evolution, respectively. Since the spherical W-S distribution has no geometric effect, we use the spherical W-S distribution to understand the role of normal fluctuation effects on our analytical results by setting $\beta_n=0$ in Eq.~(\ref{Eq:4}). We apply the kinematic cuts 0.2 GeV $\leq p_T \leq$ 3 GeV and $|\eta|\leq 0.8$. Even with the inclusion of fluctuations, the $\text{NSC}(3,\ 2)$ maintains the same sign as the analytical result, remains distinguishable between the $\alpha+^{16}$O and $5\alpha$ configurations in the $0$-$5\%$ centrality range, and the initial-state discernibility is preserved in the final-state observable. While the initial-state $\rho_3$ of 5$\alpha$ configuration agrees in sign with our analytical expectation, final-state interactions ultimately reverse its sign from positive to negative. Consequently, the $\rho_3$ becomes indistinguishable between the $\alpha+^{16}$O and $5\alpha$ configurations in the final state. Unexpectedly, the initial-state $\rho_2$ for the diffuse $5\alpha$ configuration exhibits a sign inconsistent with analytical results, due to initial-state fluctuations. This very discrepancy enables the $\rho_2$ to cleanly distinguish $\alpha+^{16}$O from $5\alpha$ configuration within 0-10\% centrality, and this diagnostic power is preserved in the final-state observable. Simultaneously, we check the effect of state mixing. With the range of structural parameters given for the $5\alpha$ and $\alpha+^{16}$O configurations in Fig.~\ref{Fig:config}(c), we randomly select the projectile and target nuclei within this range. The results, shown by the open markers in Fig.~\ref{Fig:corr}(e), indicate that the outcome of state mixing is consistent with the results from the average single Brink configuration ($b=1.61 \text{ fm}$ and $\sqrt{\langle r^2\rangle} \sim 3.0 \text{ fm}$) and falls well within the systematic uncertainty range. Based on the final-state results of the hydrodynamic framework, we conclude that $\text{NSC}(3,\ 2)$ and $\rho_2(v_{2}^{2},\ \delta [p_{\mathrm{T}}])$ are robust observables for distinguishing between the $\alpha+^{16}\text{O}$ and $5\alpha$ configurations.

Furthermore, superior to collider experiments, the LHCb fixed-target experiments directly imprint the structure of individual $^{20}$Ne nucleus in ultracentral Pb+Ne collisions at $\sqrt{s_{NN}}=70.9\ \mathrm{GeV}$~\cite{LHCb:2025ixz}, enhancing cluster-structure sensitivity through suppressed nucleon-position fluctuations~\cite{Giacalone:2024ixe}. Our correlation observables can directly apply to fixed-target experiments. The highly consistent $D_1$-$D_2$ patterns between Pb+Ne (in the Supplementary Material) and Ne+Ne collisions provide complementary tests of our predictions. 

\paragraph*{Conclusion and Outlook.}

We establish a unique method to detect complicated cluster structures through heavy-ion collisions and directly connect the $\alpha$-cluster wave function to the initial state of the QGP evolution. With the Brink cluster wave function as input and event-by-event simulations using the hydrodynamic framework, we demonstrate that symmetric cumulants NSC$(3,\ 2)$ and Pearson coefficients $\rho_2(v_{2}^{2},\ \delta [p_{\mathrm{T}}])$ exhibit a characteristic sign inversion between the $\alpha+^{16}$O and $5\alpha$ configurations. These robust sign discriminations directly connect initial-state geometry effect to measurable flow patterns, enabling unambiguous identification of $\alpha$-clustering and quantitative differentiation of complex structures. Our theoretical predictions will be tested by the Ne+Ne collisions at $\sqrt{s_{NN}}=5.36\ \mathrm{TeV}$ and fixed-target Pb+Ne collisions at $\sqrt{s_{NN}}=70.9\ \mathrm{GeV}$ at the LHC, enabling the cross-energy probes of the cluster configurations in $^{20}$Ne. Beyond nuclear physics, these observables also have potential applications in topological materials~\cite{doi:10.1021/acs.chemrev.2c00800} and Coulomb explosion imaging~\cite{doi:10.1126/science.244.4903.426,Lam:2024ysj}, where similar many-body geometric symmetry emerges.

\paragraph*{Acknowledgment.}
This work is supported by the National Natural Science Foundation of China under Grants No. 12325507, No. 12547102, No. 12147101, No. 12175042, and No. 123B1011, the National Key Research and Development Program of China under Grant No. 2022YFA1604900, 2023YFA1606701, and the Guangdong Major Project of Basic and Applied Basic Research under Grant No. 2020B0301030008. 

\vspace{-1.5\baselineskip}

\bibliography{refs}

\newpage
\clearpage
\onecolumngrid
\begin{center}

    {\large \bf Supplemental materials for ``Identifying $\alpha$-cluster configurations in $^{20}$Ne via \\[2pt] ultracentral Ne+Ne Collisions''} \\[12pt]

    { 
        Pei Li,$^{1, 2}$ 
        Bo Zhou,$^{1, 2, \color{blue}*}$ and 
        Guo-Liang Ma$^{1, 2, \color{blue}\dagger}$
    } \\[3pt]

    {\small \it $^{\it{1}}$Key Laboratory of Nuclear Physics and Ion-beam Application (MOE),} \\[-1pt]
    {\small \it Institute of Modern Physics, Fudan University, Shanghai 200433, China} \\[-1pt]
    {\small \it $^{\it{2}}$Shanghai Research Center for Theoretical Nuclear Physics,} \\[-1pt]
    {\small \it NSFC and Fudan University, Shanghai 200438, China}
\end{center}

\section*{Probes for cluster structure}\label{Probes}
We typically use the eccentricity $\varepsilon_{n}$ in the center-of-mass frame to describe the initial geometrical conditions of the collision via 2D multi-pole expansion~\cite{Joslin10101983,PhysRevC.83.064904,Moreland:2014oya}. The eccentricity vectors are defined as
\begin{align}
\varepsilon_{n}e^{in\Phi}=-\frac{\int r_{\perp}^{n}e^{in\phi}\rho(r_{\perp},\phi)\mathrm{d}^{2}r}{\int r_{\perp}^{n}\rho(r_{\perp},\phi)\mathrm{d}^{2}r},\ \ \ \ \ n\geq 2.
\label{Eq:10}
\end{align}
Previous studies typically employed the Wigner rotation matrix and the Spherical harmonics $Y_{l,m}(\theta,\phi)$ for the overall integration making the calculation more complex~\cite{Jia:2021tzt,Jia:2021qyu}. However, because the density distribution obtained by the Brink cluster model can be decomposed into multiple Gaussian components, it makes the integration relatively easier to calculate, for which no excessive approximations or expansions of higher-order terms are required compared to the W-S distribution.

In the previous studies with the W-S distribution~\cite{Jia:2021qyu}, the area size of the transverse plane in the 2D-transverse plane is define as $\frac{S_{\perp}^{2}}{\pi^{2}}=\langle x^{2}\rangle\langle y^{2}\rangle-\langle xy\rangle^{2}$, where $\langle.\rangle$ represents the average value within the 2D-transverse plane. However, in the cluster model, adopting this definition would significantly exaggerate the computational weight of regions with no matter distribution. Therefore, we modify the definition of the area in the 2D-transverse plane as 
\begin{align}
\frac{S_{\perp}^{2}}{\pi^{2}}=R_{\perp}^{4}=\langle x^{2}y^{2}\rangle-\langle xy\rangle^{2},
\label{Eq:12}
\end{align}
where $R_{\perp}$ is the transverse size. The first term represents the contribution from regions where matter is present, while the second term accounts for the subtraction of the contribution from the asymmetric distribution. 

The four-particle symmetric cumulants and normalized symmetric cumulants between different eccentricity components are defined as
\begin{align}
\mathrm{SC}(m,n)=\langle \varepsilon_{m}^{2}\varepsilon_{n}^{2}\rangle-\langle\varepsilon_{m}^{2}\rangle \langle\varepsilon_{n}^{2}\rangle,
\label{Eq:15}
\end{align}
\begin{align}
\mathrm{NSC}(m,\ n)=\frac{\langle \varepsilon_{m}^{2}\varepsilon_{n}^{2}\rangle-\langle\varepsilon_{m}^{2}\rangle \langle\varepsilon_{n}^{2}\rangle}{\langle\varepsilon_{m}^{2}\rangle \langle\varepsilon_{n}^{2}\rangle}.
\label{Eq:16}
\end{align}
The normalized symmetric cumulants are utilized to scale out the dependence on the single nucleon. The normalization of $\langle \varepsilon_{n}^{2} \delta d_{\perp} \rangle$ is defined as the Pearson correlation coefficient,

\begin{equation}
\begin{aligned}
\rho_n(\varepsilon_{n}^{2},\ \delta d_{\perp})&=\frac{\mathrm{cov}(\varepsilon_{n}^{2},\ \delta d_{\perp})}{\sqrt{\mathrm{Var}(\varepsilon_{n}^{2})\mathrm{Var}(\delta d_{\perp})}}
\\&=\frac{\langle \varepsilon_{n}^{2} \delta d_{\perp} \rangle}{\sqrt{(\langle \varepsilon_{n}^{4} \rangle-\langle \varepsilon_{n}^{2}\rangle^{2})(\langle(\delta d_{\perp} )^2\rangle)}},
\label{Eq:17}
\end{aligned}
\end{equation}
where the fluctuation of the transverse plane size $\delta d_{\perp}=d_{\perp}-\int d_{\perp} \mathrm{d}\Omega$. The energy/entropy ratio $E/S$ in the initial-state T\raisebox{-0.5ex}{R}ENTo model possesses properties similar to $d_{\perp}$, but the $E/S$ correlates with a more fundamental temperature distribution, allowing it to better reproduce the experimental observables~\cite{Giacalone:2020dln}.
\section*{The nucleon density distribution}

        \begin{figure*}[htb]
		\centering
		\subfigure
		{
			\begin{minipage}[b]{.41\linewidth}
				\centering
				\includegraphics[scale=0.7]{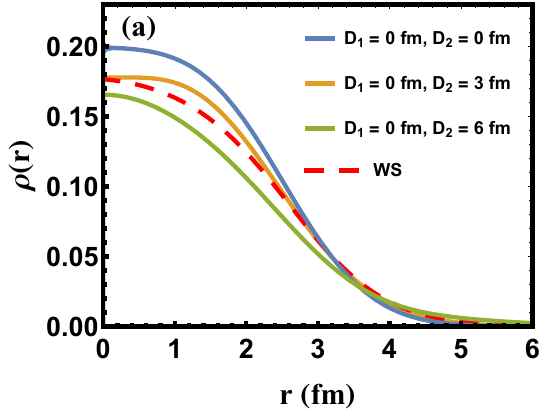}
			\end{minipage}
		}
        \subfigure
		{
			\begin{minipage}[b]{.41\linewidth}
				\centering
				\includegraphics[scale=0.7]{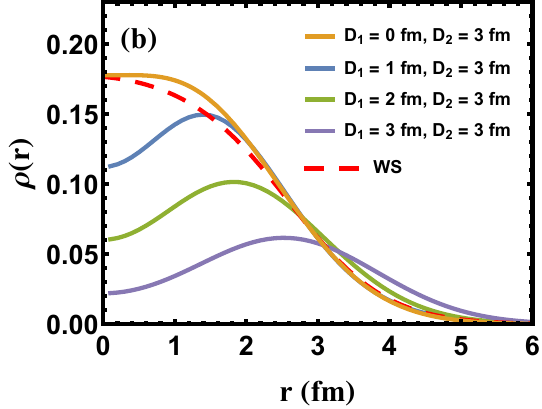}
			\end{minipage}
		}
        
		\caption{(Color online) The single nucleon radial density distribution with $b=1.76$ fm for different settings of $D_1$ and $D_2$. (a): only dependent on $D_2$ for $\alpha$+$^{16}$O configuration, (b): only dependent on $D_1$. The red dashed line represents the W-S density distribution and satisfies the experimental rms value.}
	\label{Fig:density}
	\end{figure*}

        \begin{figure*}[htb]
		\centering
		\subfigure
		{
			\begin{minipage}[b]{.31\linewidth}
				\centering
				\includegraphics[scale=0.38]{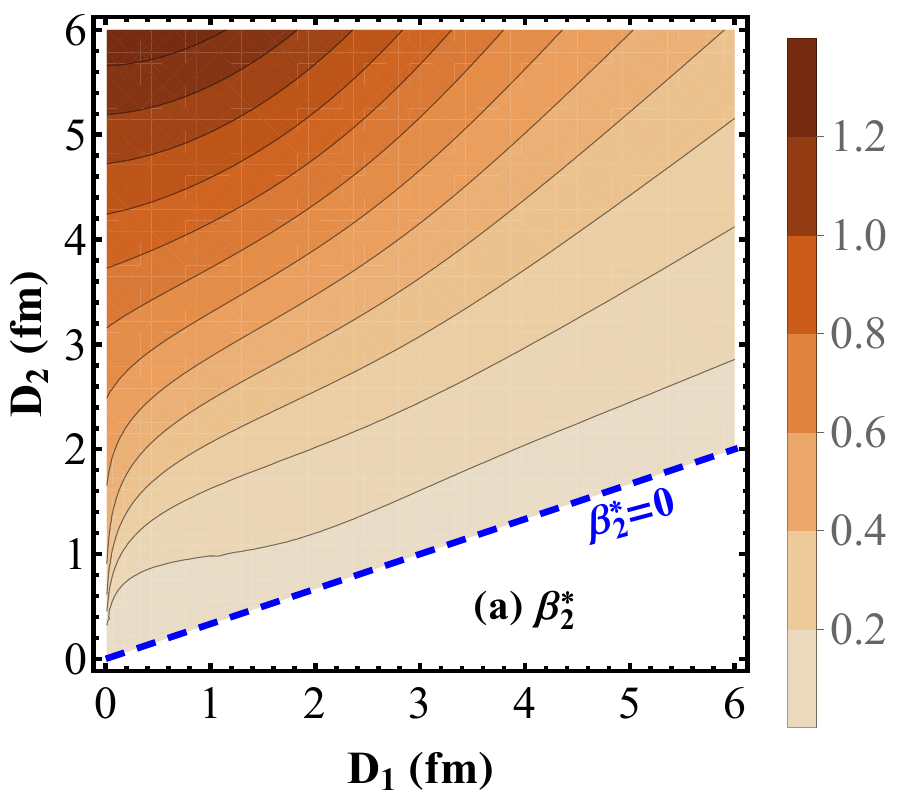}
			\end{minipage}
		}
        \subfigure
		{
			\begin{minipage}[b]{.31\linewidth}
				\centering
				\includegraphics[scale=0.38]{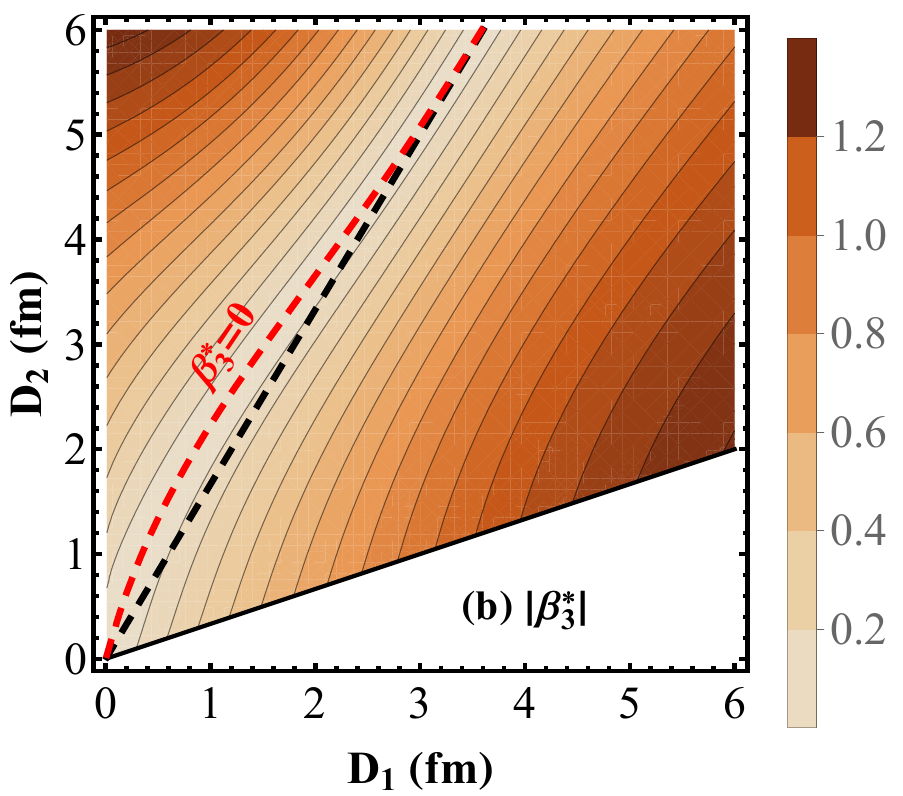}
			\end{minipage}
		}
        \subfigure
		{
			\begin{minipage}[b]{.31\linewidth}
				\centering
				\includegraphics[scale=0.38]{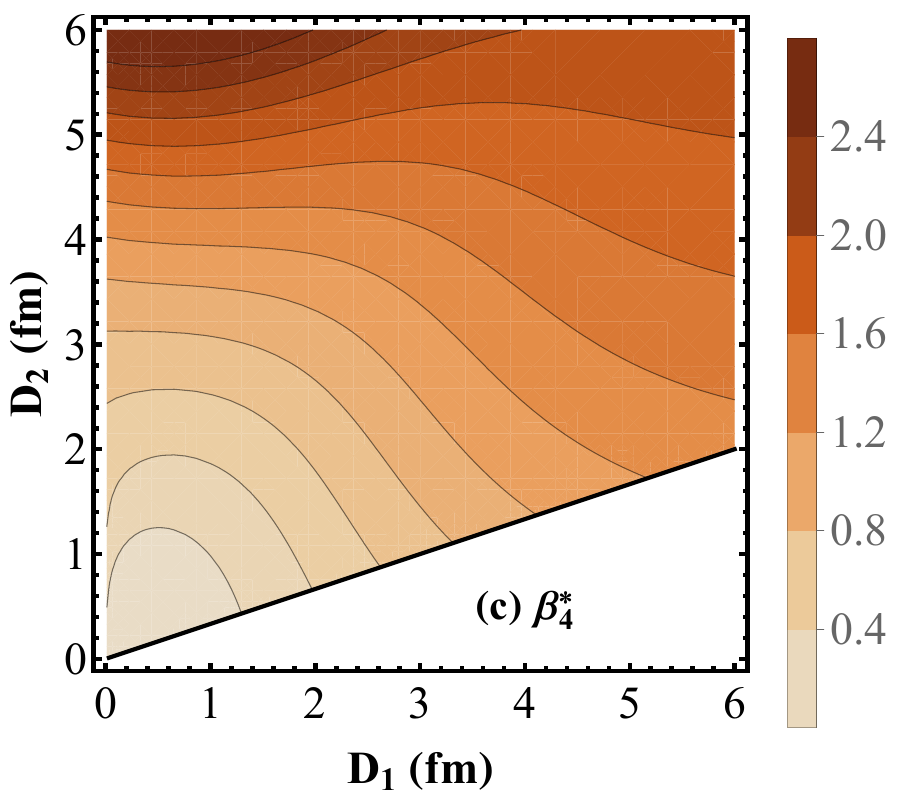}
			\end{minipage}
		}
        \subfigure
		{
            \begin{minipage}[b]{.31\linewidth}
				\centering
				\includegraphics[scale=0.38]{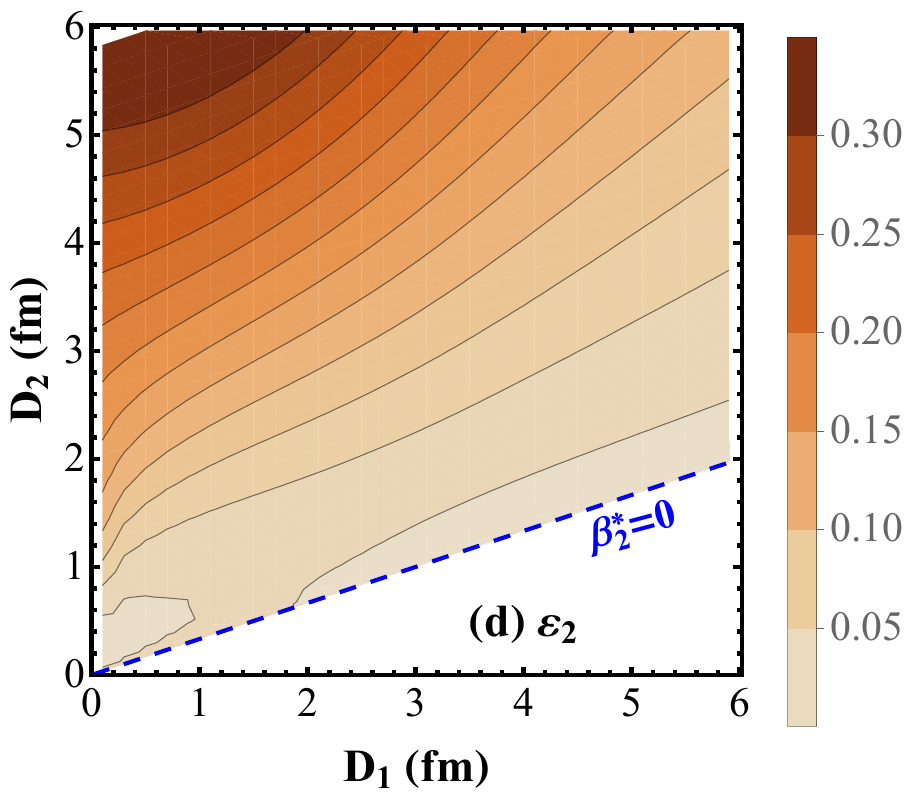}
			\end{minipage}
		}
        \subfigure
		{
			\begin{minipage}[b]{.31\linewidth}
				\centering
				\includegraphics[scale=0.38]{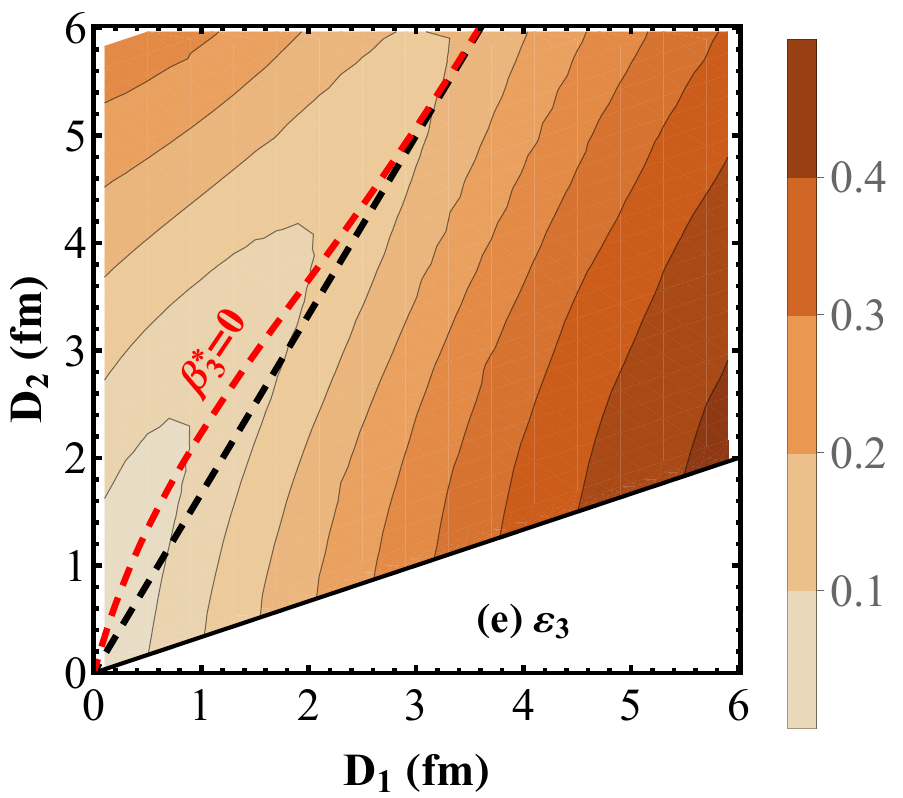}
			\end{minipage}
		}
        \subfigure
		{
			\begin{minipage}[b]{.31\linewidth}
				\centering
				\includegraphics[scale=0.38]{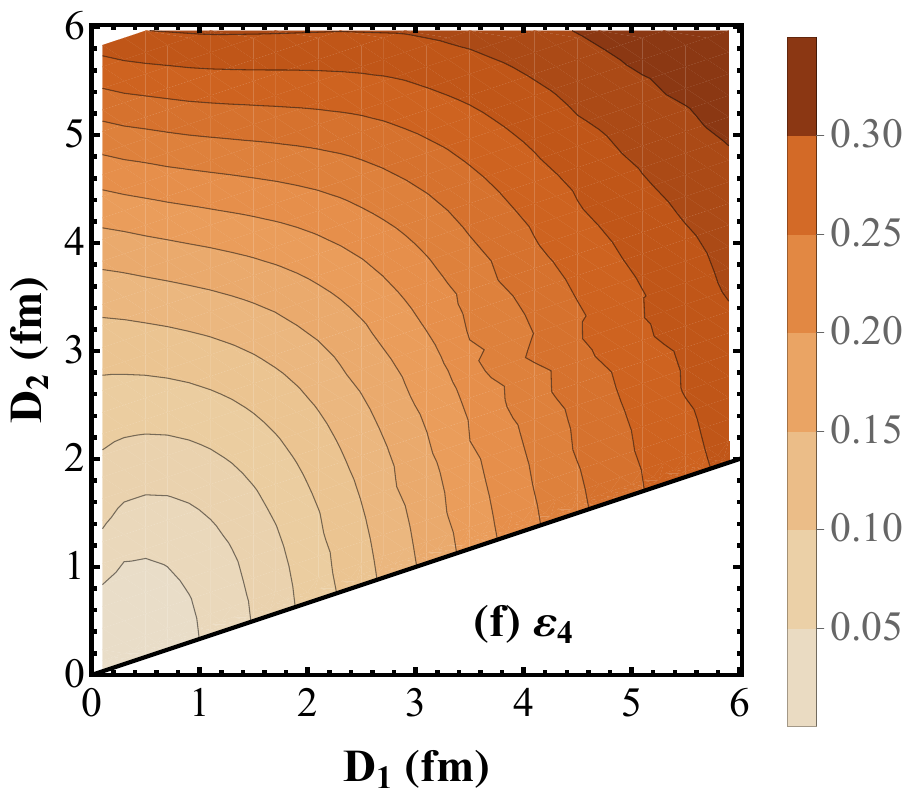}
			\end{minipage}
		}%
		\caption{(Color online) The analytical results on the dependence of the deformation parameters (a) $\beta_{2}^{*}$, (b) $\beta_{3}^{*}$, (c) $\beta_{4}^{*}$, and different orders of eccentricities (d) $\varepsilon_{2}$, (e) $\varepsilon_{3}$, (f) $\varepsilon_{4}$ on the parameters $D_1$ and $D_2$. The blue dashed lines in (a) and (d) indicate that $\beta_{2}^{*}=0$. The red dashed lines in (b) and (e) indicate that $\beta_{3}^{*}=0$. The black dashed lines in (b) and (e) present the region satisfying $k=D_{2}/D_{1}=5/3$.}
	\label{Fig:beta}
	\end{figure*}
    
Figures~\ref{Fig:density}(a) and (b) show the single nucleon radial density distribution for $^{20}$Ne with $D_1$ and $D_2$ dependence, respectively. In Fig.~\ref{Fig:density}(a), the saturation density decreases for $\alpha$+$^{16}$O configuration as $D_2$ increases. Particularly, when $D_1\rightarrow0$ and $D_2=3$ fm, the saturation density is closest to that of the W-S distribution. The density distribution does not show a significant decrease in the central region in Fig.~\ref{Fig:density}(a), while in Fig.~\ref{Fig:density}(b), the density in the central region decreases more significantly as $D_1$ increases. This indicates that the parameter $D_1$ controls the compactness of the alpha clusters within the 5$\alpha$ cluster structure. 
    
\section*{Deformation and eccentricity}
Figures~\ref{Fig:beta}(a)$-$(c) illustrate the dependence of the deformation parameters $\beta_{n}^{*}$ ($n$=2, 3, and 4) on $D_1$ and $D_2$. The images for different $\beta_{n}^{*}$ are not the same. The $\beta_{2}^{*}$ increases as $D_2$ increases, but decreases as $D_1$ increases, and becomes exactly zero with the tetrahedron configuration at the blue dashed line satisfying $k=1/3$. The $\beta_{3}^{*}$ increases as $D_1$ or $D_2$ increases and is exactly zero at the red dashed lines, which does not strictly coincide with a bi-pyramidal structure with $\mathcal{D}_{3h}$ symmetry at the black dashed line satisfying $k=5/3$ due to the Pauli exclusion effect. Figs.~\ref{Fig:beta}(d)$-$(f) present that the dependence of different eccentricity components $\varepsilon_{n}$ ($n$=2, 3, and 4) on $D_1$ and $D_2$. There are consistent proportional correspondence between $\beta_{n}^{*}$ and $\varepsilon_{n}$, $\varepsilon_{n}^2 \propto \beta_{n}^{*2}$, which are similar to the proportional relationships $\varepsilon_{n}^2 \propto \beta_{n}^{2}$ as mentioned in Ref.~\cite{Jia:2021tzt}. However, when $\beta_{3}^{*}=0$, the $\varepsilon_{3}$ is not strictly zero due to the geometric fluctuations of the overlapping region, as shown at the red dashed line in Fig.~\ref{Fig:beta}(e). This effect is reflected again in the discussion on different cumulants.

\section*{The inverse transverse area calculation}
We investigate the difference between the new definition and old definition of $S_{\perp}^{2}/\pi^{2}$ with theoretical methods using W-S distribution and the Wigner rotation matrix as shown in Ref.~\cite{Jia:2021qyu},
\begin{align}
\begin{aligned}
\frac{S_{\perp new}^{2}}{\pi^{2}}&=\langle x^{2}y^{2}\rangle-\langle xy\rangle^{2},
\label{Eq:A1}
\end{aligned}
\end{align}

\begin{align}
\begin{aligned}
\frac{S_{\perp old}^{2}}{\pi^{2}}&=\langle x^{2}\rangle\langle y^{2}\rangle-\langle xy\rangle^{2}.
\label{Eq:A2}
\end{aligned}
\end{align}
We can apply spherical harmonics $Y_n^n=\sqrt{\frac{(2n+1)!!}{4\pi(2n)!!}}\sin^n(\theta)e^{in\phi}$ to define the boundary of the W-S distribution, which serves as the upper limit for the radial integral, $r\in[0,R_0(1+$ $\sum_{l,m}\beta_l\alpha_{l,m}Y_{l,m})]$. For convenience in comparison, we disregard the second term in Eqs.~(\ref{Eq:A1}) and ~(\ref{Eq:A2}) with the effects of the misalignment of the symmetry axis with the coordinate axis. In spherical harmonics, we can use the relation,

\begin{align} 
\begin{aligned}
\sin^{4}{(\theta)}\cos^{2}{(\phi)}\sin^{2}{(\phi)}&= \frac{1}{9}-\frac{4}{9}\sqrt{\frac{\pi}{5}}Y_2^0+\frac{4\pi}{45}(Y_2^0)^2-\frac{2\pi}{15}\left((Y_2^2)^2+2Y_2^2Y_2^{-2}+(Y_2^{-2})^2\right)
\\&=\frac{1}{15}-\frac{4}{21}\sqrt{\frac{\pi}{5}}Y_2^0+\frac{2\sqrt{\pi}}{105}Y_4^0-\frac{1}{3}\sqrt{\frac{2\pi}{35}}(Y_4^4+Y_4^{-4}),
\end{aligned}
\label{Eq:A3}
\end{align}
where the second equation uses the expansion formula for spherical harmonics $Y_l^m$ with Wigner 3j symbol $\begin{pmatrix}l_1&l_2&l_3\\m_1&m_2&m_3\end{pmatrix}$,
\begin{align} 
\begin{aligned}
Y_{l_1m_1}(\theta,\phi)Y_{l_2m_2}(\theta,\phi)=\sum_{l_3=0}^\infty\sum_{m=-l_3}^lC_{l_3,m_3}Y_{l_3,m_3}(\theta,\phi),
\end{aligned}
\label{Eq:A4}
\end{align}
\begin{align} 
\begin{aligned}
C_{l_3,m_3}&=\int Y_{l_1m_1}(\theta,\phi)Y_{l_2m_2}(\theta,\phi)Y_{l_3m_3}(\theta,\phi)\mathrm{d}\Omega
\\&=\sqrt{\frac{(2l_1+1)(2l_2+1)(2l_3+1)}{4\pi}}\begin{pmatrix}l_1&l_2&l_3\\0&0&0\end{pmatrix}\begin{pmatrix}l_1&l_2&l_3\\m_1&m_2&m_3\end{pmatrix}.
\end{aligned}
\label{Eq:A5}
\end{align}

Through Eq.~(\ref{Eq:A3}), $\langle x^{2}y^{2} \rangle$ can be formally expressed in the ultracentral collisions as

\begin{align}
\begin{aligned}
&S_{\perp new}^{2}/\pi^{2}=\langle x^{2}y^{2} \rangle
=\frac{\int r^{4}\sin^{4}{(\theta)}\cos^{2}{(\phi)}\sin^{2}{(\phi)}\rho(r,\theta,\phi)\mathrm{d}^{3}r}{\int\rho(r,\theta,\phi)d^{3}r}
\\&=\frac{3R_0^4}{28\pi}\int\left(1+\sum_{l,m}\beta_l\alpha_{l,m}Y_l^m\right)^7\left[\frac{1}{15}-\frac{4}{21}\sqrt{\frac{\pi}{5}}Y_2^0+\frac{2\sqrt{\pi}}{105}Y_4^0-\frac{1}{3}\sqrt{\frac{2\pi}{35}}(Y_4^4+Y_4^{-4})\right]\sin{(\theta)}\mathrm{d}\theta \mathrm{d}\phi
\\& \approx\frac{R_0^4}{35}+\frac{3R_0^4}{28\pi} \left[ -\frac{4}{3}\sqrt{\frac{\pi}{5}}\sum_{m}\beta_2\alpha_{2,m}D_{0,m}^2+\frac{7}{5}\sum_{l_1,m_1,l_2,m_2}\beta_{l}^2\alpha_{l,m}^2+\frac{2\sqrt{\pi}}{15}\sum_{m}\beta_4\alpha_{4,m}[D_{0,m}^4-\sqrt{\frac{35}{2}}(D_{4,m}^4+D_{-4,m}^4)]
\right.\\&\left. -2\sqrt{\frac{\pi}{5}}\sum_{l_1,m_1,l_2,m_2}\beta_{l_1}\beta_{l_2}\alpha_{l_1,m_1}\alpha_{l_2,m_2}D_{0,m}^2\sqrt{\frac{5(2l_1+1)(2l_2+1)}{4\pi}}\begin{pmatrix}l_1&l_2&2\\0&0&0\end{pmatrix}\begin{pmatrix}l_1&l_2&2\\m_1&m_2&m\end{pmatrix}
\right.\\&\left.+\frac{2\sqrt{\pi}}{5}\sum_{l_1,m_1,l_2,m_2}\beta_{l_1}\beta_{l_2}\alpha_{l_1,m_1}\alpha_{l_2,m_2}[D_{0,m}^4-\sqrt{\frac{35}{2}}(D_{4,m}^4+D_{-4,m}^4)]\sqrt{\frac{9(2l_1+1)(2l_2+1)}{4\pi}}\begin{pmatrix}l_1&l_2&4\\0&0&0\end{pmatrix}\begin{pmatrix}l_1&l_2&4\\m_1&m_2&m\end{pmatrix}\right],
\label{Eq:A6}
\end{aligned}
\end{align}
where the fourth approximate equality uses the expansion $(1+x)^a = 1+ax+\frac{a(a-1)}{2}x^2+O(x^3)$ and $D_{m,m^{\prime}}^n$ is the Wigner rotation matrix. Thus, we can obtain the inverse transverse area with the new definition,

\begin{align}
\begin{aligned}
&S_{\perp new}^{-1}=\frac{\sqrt{35}}{\pi R_0^2}\left[1+\sqrt{\frac{5}{4\pi}}\sum_{m}\beta_2\alpha_{2,m}D_{0,m}^2+\frac{7}{5}\sum_{l_1,m_1,l_2,m_2}\beta_{l}^2\alpha_{l,m}^2-\frac{1}{4\sqrt{\pi}}\sum_{m}\beta_4\alpha_{4,m}[D_{0,m}^4-\sqrt{\frac{35}{2}}(D_{4,m}^4+D_{-4,m}^4)]
\right.\\&\left.+\frac{3}{4}\sqrt{\frac{5}{\pi}}\sum_{l_1,m_1,l_2,m_2}\beta_{l_1}\beta_{l_2}\alpha_{l_1,m_1}\alpha_{l_2,m_2}D_{0,m}^2\sqrt{\frac{5(2l_1+1)(2l_2+1)}{4\pi}}\begin{pmatrix}l_1&l_2&2\\0&0&0\end{pmatrix}\begin{pmatrix}l_1&l_2&2\\m_1&m_2&m\end{pmatrix}
\right.\\&\left.-\frac{3}{4\sqrt{\pi}}\sum_{l_1,m_1,l_2,m_2}\beta_{l_1}\beta_{l_2}\alpha_{l_1,m_1}\alpha_{l_2,m_2}[D_{0,m}^4-\sqrt{\frac{35}{2}}(D_{4,m}^4+D_{-4,m}^4)]\sqrt{\frac{9(2l_1+1)(2l_2+1)}{4\pi}}\begin{pmatrix}l_1&l_2&4\\0&0&0\end{pmatrix}\begin{pmatrix}l_1&l_2&4\\m_1&m_2&m\end{pmatrix}\right],
\label{Eq:A7}
\end{aligned}
\end{align}
and the deviation of the inverse transverse area can be expressed as 
\begin{align}
\begin{aligned}
&\delta S_{\perp new}^{-1}=\frac{\sqrt{35}}{\pi R_0^2}\left[\sqrt{\frac{5}{4\pi}}\sum_{m}\beta_2\alpha_{2,m}\delta D_{0,m}^2-\frac{1}{4\sqrt{\pi}}\sum_{m}\beta_4\alpha_{4,m}\delta[ D_{0,m}^4-\sqrt{\frac{35}{2}}(\delta D_{4,m}^4+\delta D_{-4,m}^4)]
\right.\\&\left.+\frac{3}{4}\sqrt{\frac{5}{\pi}}\sum_{l_1,m_1,l_2,m_2}\beta_{l_1}\beta_{l_2}\alpha_{l_1,m_1}\alpha_{l_2,m_2}\delta D_{0,m}^2\sqrt{\frac{5(2l_1+1)(2l_2+1)}
{4\pi}}\begin{pmatrix}l_1&l_2&2\\0&0&0\end{pmatrix}\begin{pmatrix}l_1&l_2&2\\m_1&m_2&m\end{pmatrix}
\right.\\&\left.-\frac{3}{4\sqrt{\pi}}\sum_{l_1,m_1,l_2,m_2}\beta_{l_1}\beta_{l_2}\alpha_{l_1,m_1}\alpha_{l_2,m_2}\delta [D_{0,m}^4-\sqrt{\frac{35}{2}}(D_{4,m}^4+D_{-4,m}^4)]\sqrt{\frac{9(2l_1+1)(2l_2+1)}{4\pi}}\begin{pmatrix}l_1&l_2&4\\0&0&0\end{pmatrix}\begin{pmatrix}l_1&l_2&4\\m_1&m_2&m\end{pmatrix}\right].
\label{Eq:A8}
\end{aligned}
\end{align}

We keep higher-order $\beta_2$ terms and neglect higher-order $\beta_4$ terms in the expansion of the numerator in Eq.~(\ref{Eq:A7}) and Eq.~(\ref{Eq:A8}). With the old definition in Eq.~(\ref{Eq:A2}), Ref.~\cite{Wang:2024ulq} has presented the result of the inverse transverse area $\langle(\delta S_{\perp old}^{-1})^2\rangle\approx \frac{0.25}{\pi^2R_0^4}(7.954\beta_{2}^{2}- 4.301\beta_{2}^{3}+5.352\beta_{2}\beta_{3}^{2}+O(\beta_{n}^{4}))$. We can calculate $\langle(\delta S_{\perp new}^{-1})^2\rangle$ by Eq.~(\ref{Eq:A8}),

\begin{align}
\begin{aligned}
&\langle(\delta S_{\perp new}^{-1})^2\rangle=\frac{0.5}{\pi^2R_0^4}(5.57\beta_{2}^{2}+1.506\beta_{2}^{3}+1.406\beta_{2}\beta_{3}^{2}+2.572\beta_{3}^2\beta_{4}+8.081\beta_{2}^2\beta_{4}+O(\beta_{n}^{4})).
\label{Eq:A9}
\end{aligned}
\end{align}

\begin{figure*}[tb]
		\centering
        \subfigure
		{
			\begin{minipage}[b]{0.4\linewidth}
				\centering
				\includegraphics[scale=0.4]{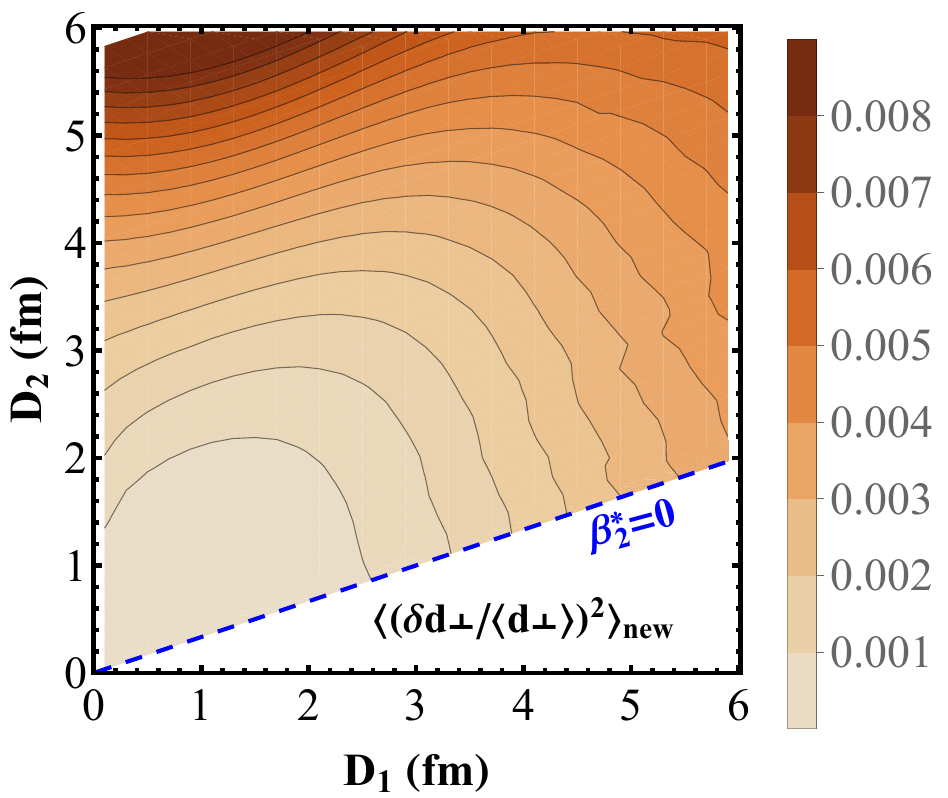}
			\end{minipage}
		}%
        \subfigure
		{
			\begin{minipage}[b]{0.4\linewidth}
				\centering
				\includegraphics[scale=0.4]{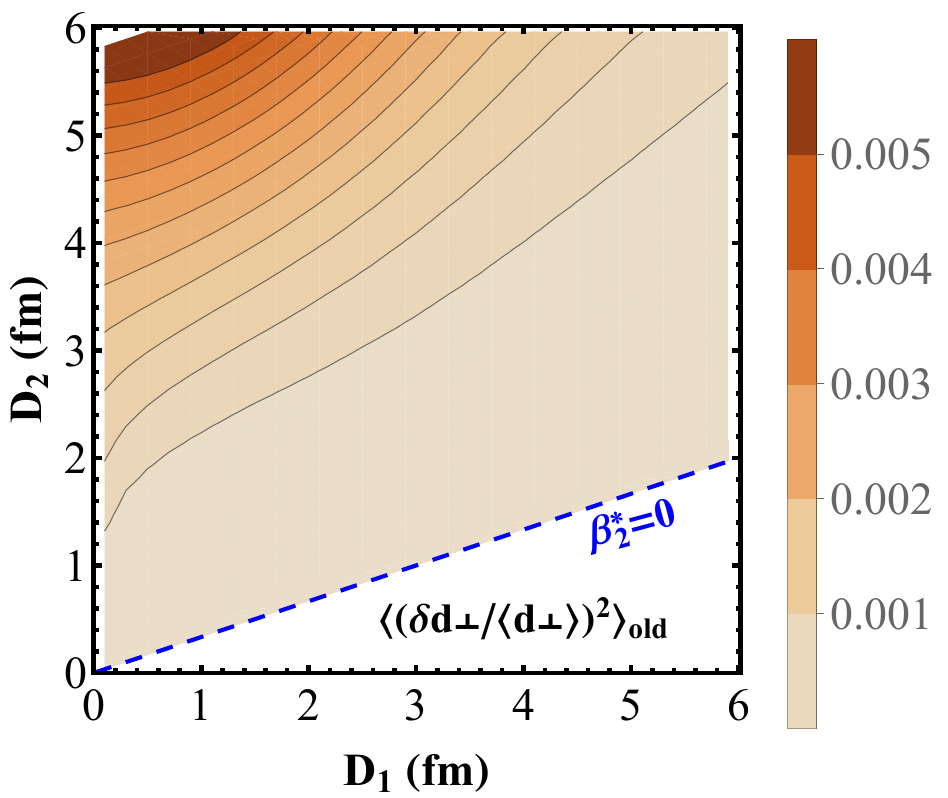}
			\end{minipage}
		}%
		\caption{(Color online) The analytical results on the dependence of the fluctuation of the inverse transverse size $\langle(\delta d_{\perp}/\langle d_{\perp} \rangle)^2\rangle$ with new definition (left) and old definition (right) on the parameters $D_1$ and $D_2$. The blue dashed lines indicate that $\beta_{2}^{*}=0$.}
        
	\label{Fig:ddc2}
	\end{figure*}

According to the reliable $\alpha$-particle scattering and proton scattering experiments from Ref.~\cite{Burtebayev:2018woh,deSwiniarski:1976rcl}, the hexadecapole deformation is found to be $\beta_4 = 0.17 \pm 0.06$. Using the deformation parameters $\beta_2$ and $\beta_3$ from Ref.~[57], the introduction of $\beta_4$ results in an enhancement of $\langle (\delta S^{-1}_{\perp \text{new}})^2\rangle$ by $(21 \pm 7)\%$. With the relation $\frac{\delta d_{\perp}}{\langle d_{\perp} \rangle}=\frac{1}{2}\frac{\delta S_{\perp}^{-1}}{\langle S_{\perp}^{-1}\rangle}$, the dependence of $\langle(\delta d_{\perp}/\langle d_{\perp}\rangle )^{2}\rangle$ on $D_1$ and $D_2$ by new definition and old definition are shown in Fig.~\ref{Fig:ddc2}. The old definition can perfectly match the proportional relationship $\langle(\delta d_{\perp}/ \langle d_{\perp} \rangle)^{2}\rangle \propto \beta_{2}^{2}$, while the properties of new definition does not match the relation in the region with small $\beta_2$ due to the inclusion of the $\beta_4$ contribution, which better reconciles the theoretical framework with reality. For the same configuration, the value of $\langle (\delta d_{\perp})^{2}\rangle$ under the new definition is larger. This indicates that the new definition significantly reduces the area erroneously estimated due to the differences between the cluster model and the W-S distribution. For the cluster model, there is typically a cavity structure at the center, where the central density slightly decreases compared to the highest point. Therefore, using the old definition where two multiplied average value does not accurately reflect such a complex structure. However, this definition is applicable to the W-S distribution, for which the density gradually decreases from the center to the edge radially without exhibiting a cavity structure.

    \begin{figure*}[tb]
		\centering
        \subfigure
		{
			\begin{minipage}[b]{0.315\linewidth}
				\centering
				\includegraphics[scale=0.38]{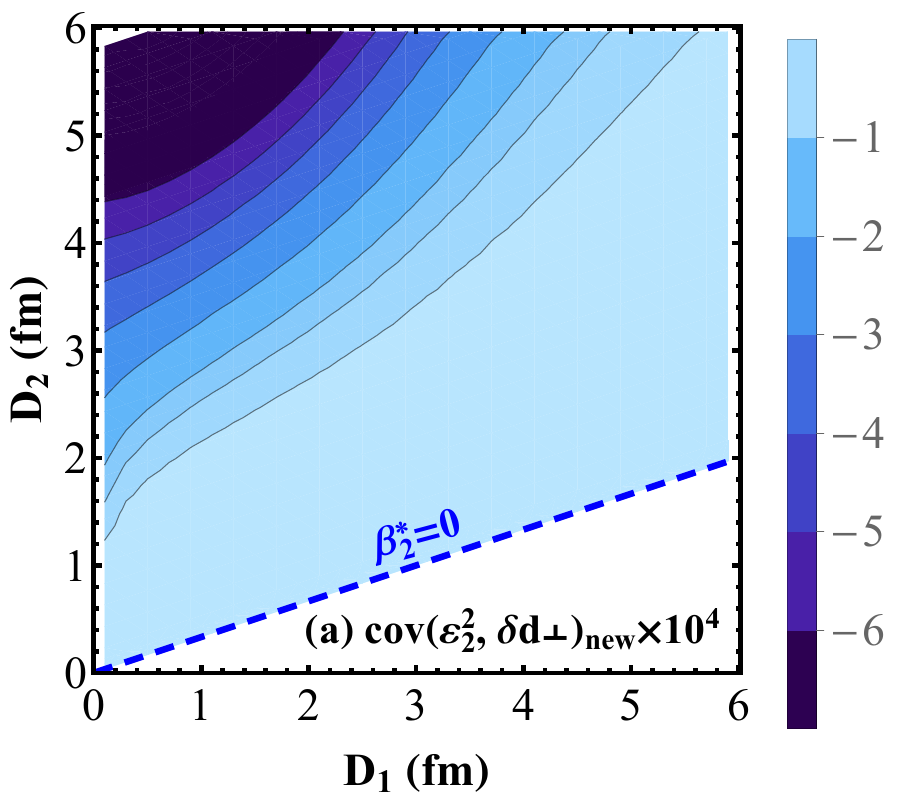}
			\end{minipage}
		}%
        \subfigure
		{
			\begin{minipage}[b]{0.315\linewidth}
				\centering
				\includegraphics[scale=0.38]{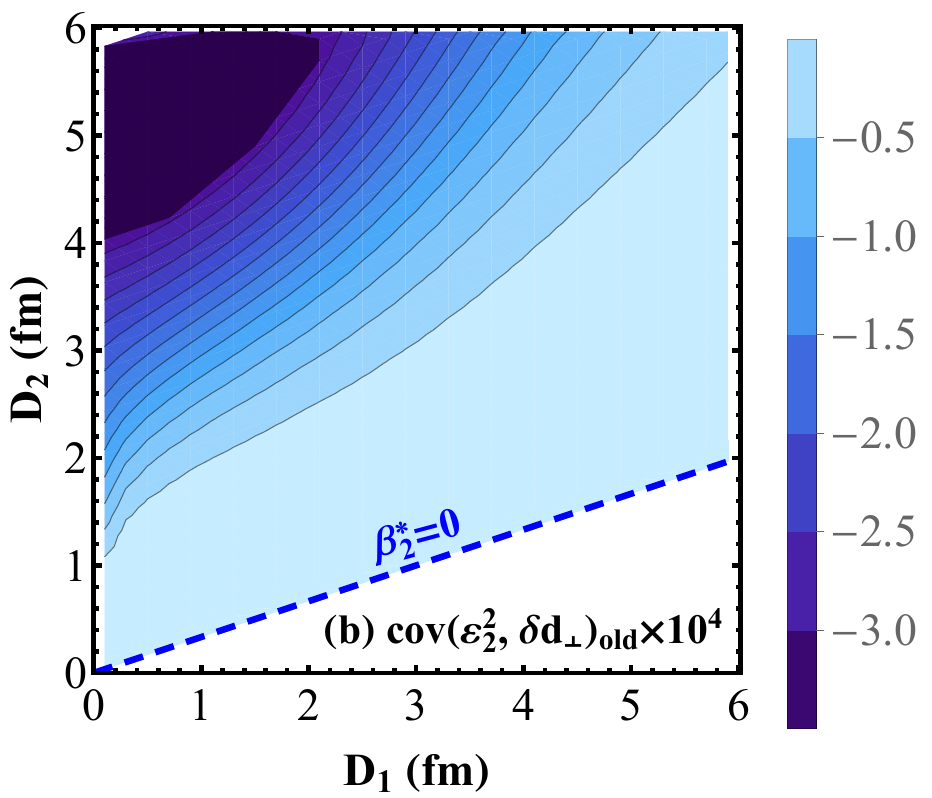}
			\end{minipage}
		}%
        \subfigure
		{
			\begin{minipage}[b]{0.315\linewidth}
				\centering
				\includegraphics[scale=0.38]{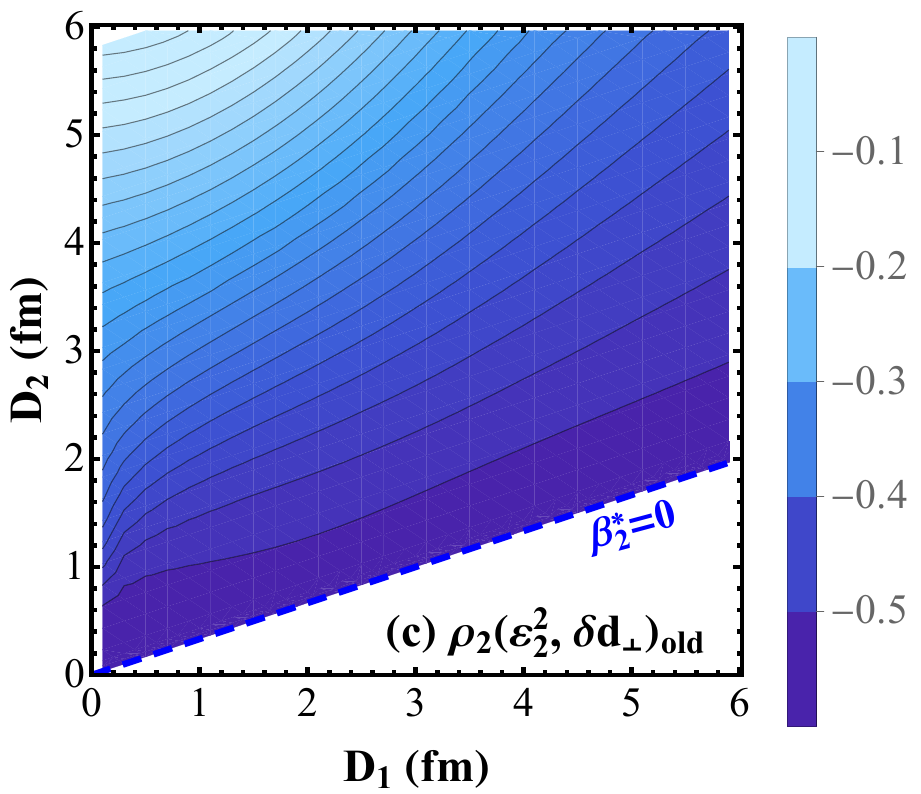}
			\end{minipage}
		}%
		\caption{(Color online) The analytical results on the dependence of the covariance $\mathrm{cov}(\varepsilon_2^2,\ \delta d_{\perp})$ (left: new definition, middle: old definition) and the Pearson correlation coefficient $\rho_2$ between $\varepsilon_2$ and $\langle d_{\perp}\rangle$ with old definition on $D_1$ and $D_2$ (right). The blue dashed lines indicate that $\beta_{2}^{*}=0$.}
        
	\label{Fig:pccold}
	\end{figure*}  
    
The old definition not only fails to correctly describe the transverse area of the cluster model but also cannot accurately capture the fluctuations of the transverse area. For example, When we calculate $\mathrm{cov}(\varepsilon_{2}^{2},\ \delta d_{\perp})$ and $\rho_2$ with the old definition, as shown in Figs.~\ref{Fig:pccold}(b) and (c), we observe that $\mathrm{cov}(\varepsilon_{2}^{2},\ \delta d_{\perp}) \rightarrow0$ but $\rho_2$ is not zero with $\beta_2^* \rightarrow0$. Moreover, the dependence of $\mathrm{cov}(\varepsilon_{2}^{2},\ \delta d_{\perp} )$ and $\rho_2$ is completely opposite, which reflects erroneous information and does not align with physical reality. For the new definition for the calculation, both $\mathrm{cov}(\varepsilon_{2}^{2},\ \delta d_{\perp} )$ (Fig.~\ref{Fig:pccold}(a)) and $\rho_2$ exhibit more consistent behavior.

\section*{Fixed-target collisions}
\begin{figure*}[ht]
		\centering
        \subfigure
		{
			\begin{minipage}[b]{0.315\linewidth}
				\centering
				\includegraphics[scale=0.38]{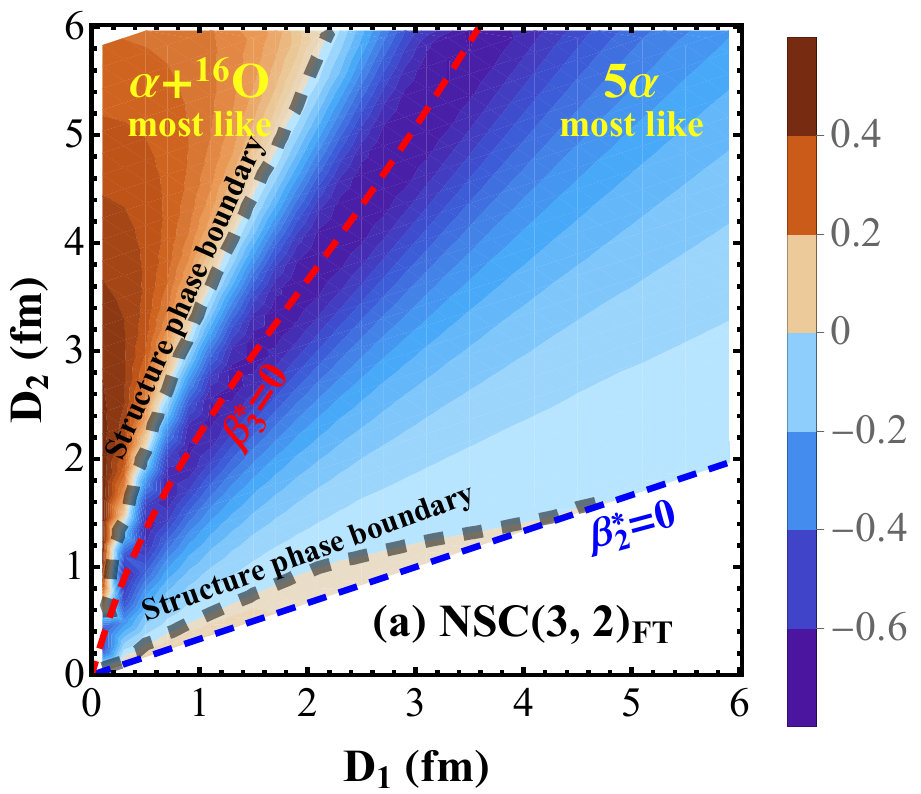}
			\end{minipage}
		}%
        \subfigure
		{
			\begin{minipage}[b]{0.315\linewidth}
				\centering
				\includegraphics[scale=0.38]{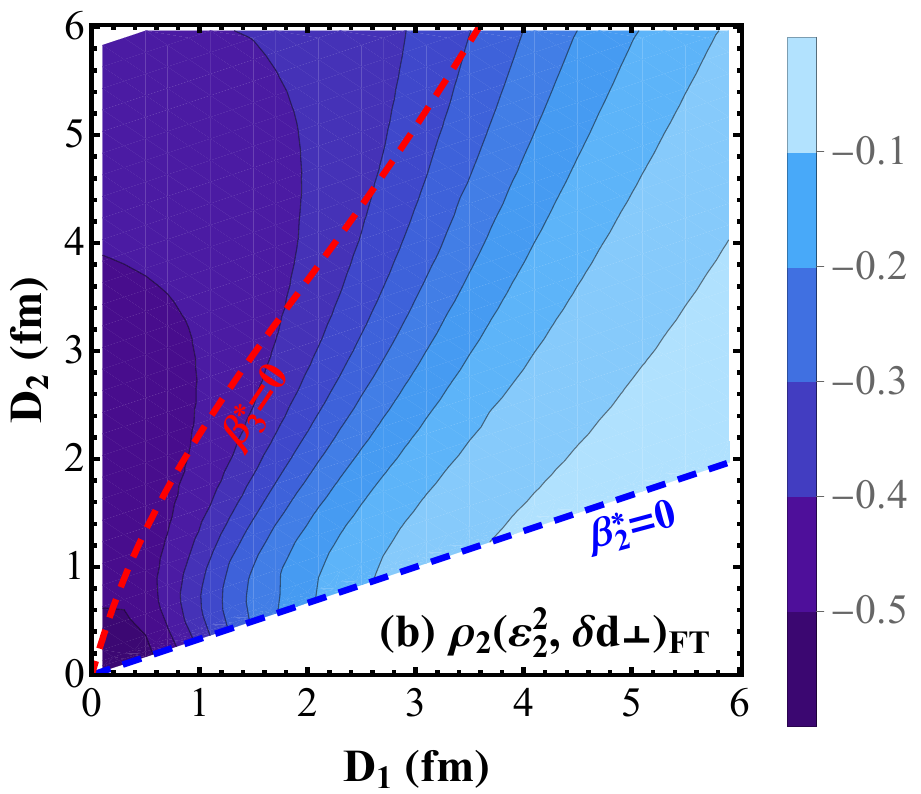}
			\end{minipage}
		}%
        \subfigure
		{
			\begin{minipage}[b]{0.315\linewidth}
				\centering
				\includegraphics[scale=0.38]{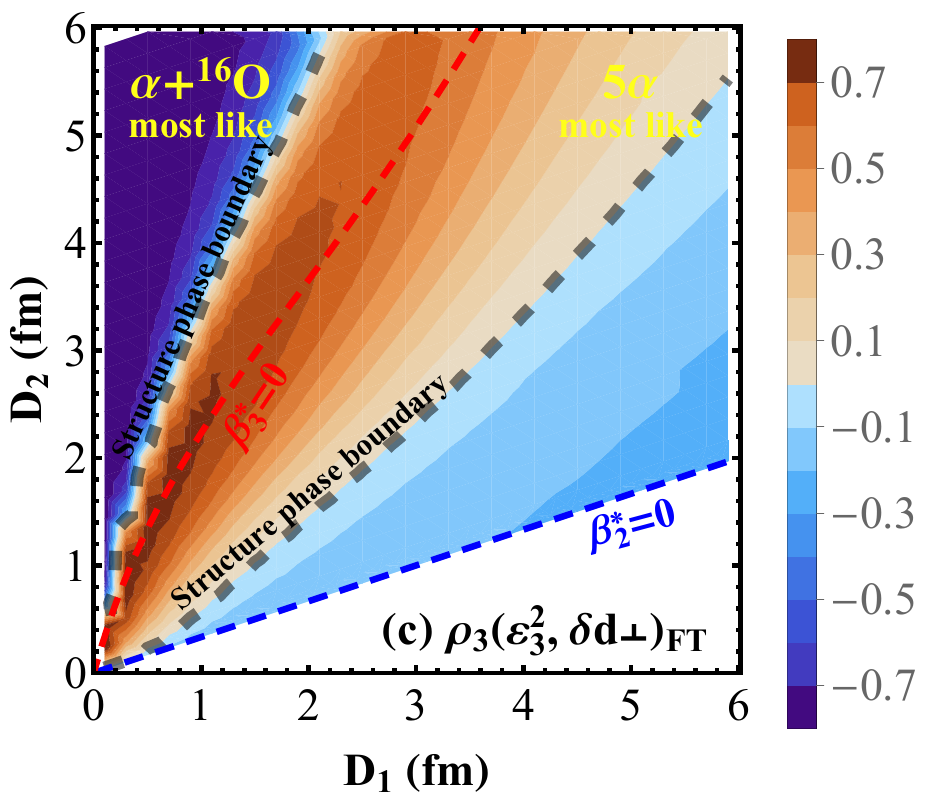}
			\end{minipage}
		}%
		\caption{(Color online) The analytical results on the dependence of the correlation observables on the parameters $D_1$ and $D_2$ with $b=1.76$ fm in fixed-target Pb+Ne collisions, (a): NSC$(3,\ 2)$, (b): $\rho_2$, (c): $\rho_3$. The blue dashed lines indicate that $\beta^{*}_2=0$. The red dashed line indicates that $\beta^{*}_3=0$. The gray thick dashed lines are the structure phase boundaries that distinguish $\alpha+^{16}$O and 5$\alpha$ configurations.}
        
	\label{Fig:FT}
	\end{figure*}

Figures~\ref{Fig:FT}(a)-(c) illustrate the analytical results of the normalized symmetric cumulants NSC$(3,\ 2)$ and Pearson correlation coefficient $\rho_2$ and $\rho_3$ in the fixed-target Pb+Ne ultracentral collisions, respectively. We achieve an approximate fixed-target effect by assuming that the rotational direction of the projectile and target nuclei are aligned. The resulting correlation results for the fixed target agree perfectly with the results of the Ne+Ne collisions. Since the fixed-target ultracentral collisions contain all structural information of $^{20}$Ne without any spectators, the enhanced strength of the correlations shows that the cluster structural effect is even more pronounced in the fixed-target experiments. The results of Pb+Ne collisions obtained from the 2D T\raisebox{-0.5ex}{R}ENTo simulations align with the Ne+Ne collisions. Moving forward, a more applicable 3D framework should be performed as additional validation, such as 3DGlauber+MUSIC+UrQMD~\cite{Giacalone:2024ixe}.

\section*{Hydrodynamic response coefficient}
\begin{figure}[htbp]
		\centering
        \subfigure
		{
			\begin{minipage}[b]{.45\linewidth}
				\centering
				\includegraphics[scale=0.62]{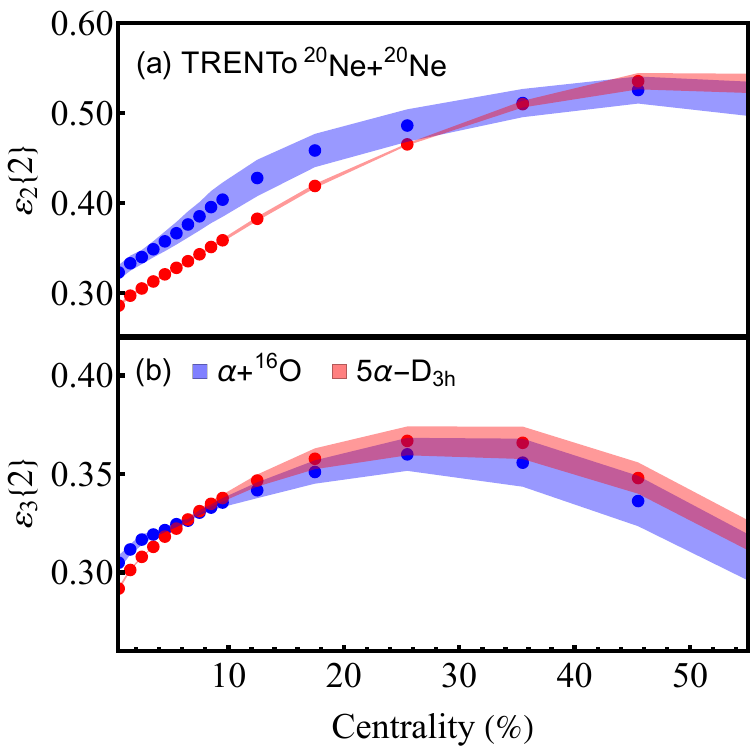}
			\end{minipage}
		}
        \subfigure
		{
			\begin{minipage}[b]{.45\linewidth}
				\centering
				\includegraphics[scale=0.62]{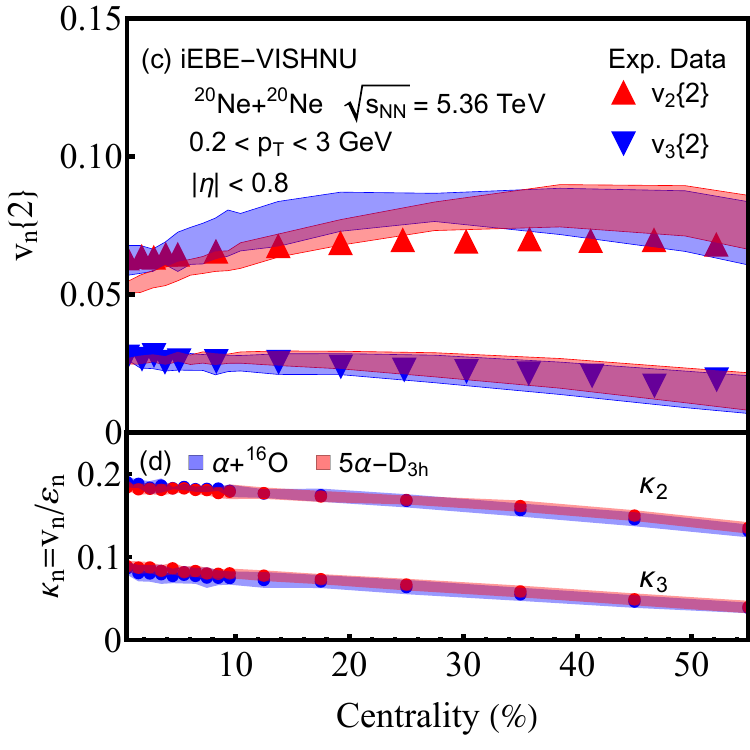}
			\end{minipage}
		}

        \caption{(Color online) The eccentricities as a function of centrality in Ne+Ne collisions, (a) $\varepsilon_2\{2\}$  and (b) $\varepsilon_3\{2\}$. (c): Charged-particle anisotropic flow coefficients $v_2\{2\}$ and $v_3\{2\}$ as a function of centrality in Ne+Ne collisions at $\sqrt{s_{NN}} = 5.36\ \mathrm{TeV}$ for the $5\alpha$ (blue band) and $\alpha+^{16}\mathrm{O}$ (red band) configurations. The bands represent systematic uncertainties. Our results are compared with ALICE data~\cite{ALICE:2025luc}. (d): Corresponding hydrodynamic response coefficients $\kappa_2$ and $\kappa_3$ as a function of centrality.}
    \label{Fig:vnkn}
    \end{figure}

Figures~\ref{Fig:vnkn}(a)-(c) show the results for the eccentricities $\varepsilon_n\{2\}$ and the collective flow $v_n\{2\}$ using the two-particle correlation method with the hydrodynamic framework of T\raisebox{-0.5ex}{R}ENTo+VISHNU+UrQMD. In the T\raisebox{-0.5ex}{R}ENTo model, the inelastic nucleon-nucleon cross-section is set to $\sigma_{NN}=70.9$ mb for Ne+Ne collisions at the LHC energy 5.36 TeV. The nucleon width is set to $w=0.5$ fm with the number of subnucleon $n_c=6$. The remaining parameters are consistent with those in Ref.~\cite{Hu:2025eid}. While the ordering of $\varepsilon_n\{2\}$ in ultra-central collisions aligns with the analytical predictions in Fig.~\ref{Fig:beta}, the inclusion of nucleon fluctuations narrows the gap between the two configurations. The relative differences $\frac{|\varepsilon_n\{2\}_I-\varepsilon_n\{2\}_{II}|}{(\varepsilon_n\{2\}_I+\varepsilon_n\{2\}_{II})/2}$ between the $5\alpha$ and $\alpha+^{16}$O configurations are 12$\%$ for $\varepsilon_2\{2\}$ and 4.4$\%$ for $\varepsilon_3\{2\}$. The $v_n\{2\}$ after the whole evolution of heavy-ion collisions can not distinguish between the two configurations, and both configurations show comparable magnitudes that are consistent with experimental data~\cite{ALICE:2025luc} using kinematic cuts 0.2 GeV $\leq p_T \leq$ 3 GeV and $|\eta|\leq 0.8$. In Fig.~\ref{Fig:vnkn}(d), furthermore, the negligible difference of the response coefficients $\kappa_n$ for both configurations at $\sqrt{s_{NN}} = 5.36$ TeV directly supports the reliability of the linear response assumption in this Letter.
    
\section*{Pearson correlation with $\delta d_\perp$}
\begin{figure}[htbp]
		\centering

        \subfigure
		{
			\begin{minipage}[b]{.9\linewidth}
				\centering
				\includegraphics[scale=0.62]{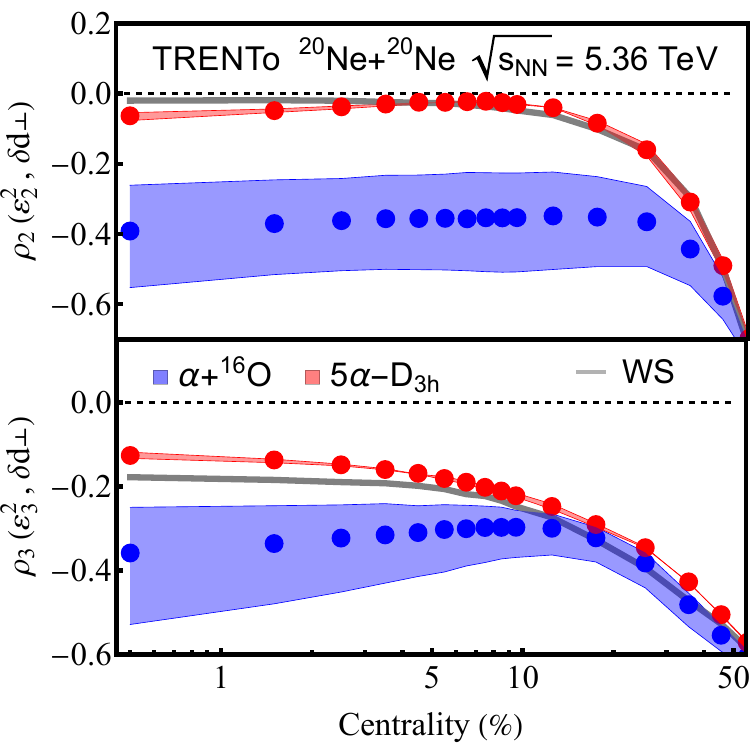}
			\end{minipage}
		}
        \subfigure

    \caption{(Color online) Pearson correlation coefficient $\rho_n(\varepsilon_n^2, \delta d_{\perp})$ as a function of centrality in Ne+Ne collisions at $\sqrt{s_{NN}} = 5.36\ \mathrm{TeV}$ for the $5\alpha$ (blue band) and $\alpha+^{16}\mathrm{O}$ (red band) configurations. Top: $n=2$. Bottom: $n=3$. The solid gray band denotes the result for the W-S distribution. The bands represent systematic uncertainties.
    }
    \label{Fig:pccddc}
    \end{figure}

To address the results of different initial-state estimators for the $v_2$-$[p_T]$ final-state correlation, Fig.~\ref{Fig:pccddc} presents the results for the initial-state correlation quantity $\rho_n(\varepsilon_n^2, \delta d_\perp)$ ($n$=2, 3) calculated using Eq.~(6). The $\rho_n(\varepsilon_n^2, \delta d_\perp)$ for the W-S distribution are negative for ultra-central collisions, contrary to the positive values using E/S as estimator in the T\raisebox{-0.5ex}{R}ENTo model as shown in Fig.~2(e) in the main text, highlighting the impact of initial-state nucleon position fluctuations. However, the differences between the $5\alpha$ and $\alpha+^{16}\text{O}$ structures in the initial-state T\raisebox{-0.5ex}{R}ENTo model can still be explained by the analytical calculation results: relative to the W-S distribution, $\rho_2$ is suppressed for both cluster-related structures (with a stronger suppression for $\alpha+^{16}\text{O}$), while two $\rho_3$ lie on opposite sides of the W-S baseline. These distinct signatures originate from the underlying geometry of the cluster configurations.

\end{document}